\documentclass{emulateapj}

\shortauthors{Terradas et al.}
\shorttitle{Effect of magnetic twist on nonlinear transverse kink oscillations}

\begin{document}

\title{Effect of magnetic twist on nonlinear transverse kink oscillations \\ of
line-tied magnetic flux tubes}
\email{jaume.terradas@uib.es}
\author{J. Terradas}

\affiliation{Departament de F\'\i sica,
Universitat de les Illes Balears (UIB), E-07122, Spain}
\affiliation{Institute of Applied Computing \& Community Code (IAC$^3$), UIB, Spain}
\author{N. Magyar and T. Van Doorsselaere}
\affiliation{Centre for Mathematical Plasma Astrophysics Department of Mathematics,\\ Celestijnenlaan
200B,  B-3001, KULeuven, Leuven, Belgium}

\begin{abstract}

Magnetic twist is thought to play an important role in many structures of the
solar atmosphere. One of the effects of twist is to modify the properties of the
eigenmodes of magnetic tubes. In the linear regime standing kink solutions are
characterized by a change in polarization of the transverse displacement along
the twisted tube. In the nonlinear regime,  magnetic twist affects the
development of shear instabilities that appear at the tube boundary when it is
oscillating laterally. These Kelvin-Helmholtz instabilities (KHI) are produced
either by the jump in the azimuthal component of the velocity at the edge of the
sharp boundary between the internal and external part of the tube, or either by
the continuous small length-scales produced by phase-mixing when there is a
smooth inhomogeneous layer.  In this work the effect of twist is consistently
investigated by solving the time-dependent problem including the process of
energy transfer to the inhomogeneous layer. It is found that twist always delays
the appearance of the shear instability  but for tubes with thin inhomogeneous
layers the effect is relatively small for moderate values of twist. On the
contrary, for tubes with thick layers, the effect of twist is much stronger. This
can have some important implications regarding observations of transverse kink
modes and the KHI itself.

\end{abstract}


\section{Introduction}

Standing transverse oscillations of cylindrical magnetic tubes are prone to
develop, in the nonlinear regime, shear instabilities at the tube boundary
\citep[e.g.,][]{terradasetal08,antolinetal2014,antolinetal2015,
magyaretal2015,magyartom2016,antolinetal2017}. The instability can take place in magnetic
tubes with either a sharp or a smooth density transition between the internal
part of the tube and the lighter external medium. 

For a sharp transition and in the linear regime, eigenmode calculations show
that the azimuthal component of the velocity is discontinuous at the tube radius
\citep[see for example,][]{goossensetal2012}. This is the origin of the
Kelvin-Helmholtz instability (KHI) found in the nonlinear regime. The velocity
shear profile is rather different from the interface equilibrium model used to
investigate this instability based on stationary and invariant flows \citep[see
for example,][]{chandra61}. For a cylindrical magnetic tube oscillating with the
standing transverse $m=1$ mode, the shear flow due to the oscillation is
sinusoidal in the azimuthal and longitudinal direction (along the axis of the
tube). Since the tube is oscillating, the flow is not stationary in time and has
the periodicity  of the kink oscillation. This complicates significantly an
analytical study of the problem even in the linear regime and numerical
simulations are in general required to address this problem.

For a smooth transition at the tube boundary the situation is even more complex.
Transverse kink oscillations are attenuated due to the energy transference to
the inhomogeneous layer by the mechanism of resonant damping
\citep[e.g.,][]{hollyang88,rudrob02,goossetal02}. \citet{sakuraietal91,
goossensetal92} showed that the perpendicular component of the velocity at the
resonant position ($r=r_{\rm A}$) has a $1/s-$singularity (where $s=r-r_{\rm
A}$) and a $\delta(s)$ contribution. Later,  \citet{goossensetal95} showed how
dissipation smooths the velocity profile (see their Fig.~1). The bulk transverse
motion of the tube is eventually transformed into highly localized shear motions
at the edge of the tube where phase-mixing takes place. In comparison with
the sharp transition configuration, there is now an additional complication, the
change in time of the shear profile in the radial direction. Decreasing spatial
scales are continuously generated at the inhomogeneous layer by the phase-mixing
process. 

\citet{heypri83} and \citet{brownpri84} were able to study  the development of
the KHI in a very simple configuration and for uncoupled Aflv\'en waves ($m=0$).
In their model, the process of resonant damping is not considered and the amount
of energy in the layer does not grow with time. However, their results can be
used as a crude guide to better understand the onset of the KHI. These authors
performed a temporally local stability analysis and found a critical time after
which the instability grows significantly within one period of oscillation. In
essence the authors solve the standard Rayleigh stability equation for a shear
flow in two different regimes. The first regime is characterized by
length-scales produced by phase-mixing that are much shorter than the typical
inhomogeneity length-scale. In this strong phase-mixing situation the shear
profile is approximated by a rapidly oscillating sinusoidal function
\citep{brownpri84}. The instability could, however, develop before the situation
of strong phase-mixing is reached. In this regime the instability at first grows
slowly, but after a critical time, the instability develops within one wave
period. The critical time depends on the ratio of the inhomogeneity scale length
to the wavelength parallel to the equilibrium magnetic field. 

The main aim of the present work is to study the stabilizing effect of an
azimuthal component of the magnetic field on the development of the KHI. In this
regard, \citet{soleretal2010} performed an analytical study considering the
linear stage of the KHI in a cylindrical magnetic tube in the presence of an
invariant azimuthal flow and one of the main conclusions was that, in principle,
a small amount of twist should be able to suppress the instability.
\citet{temurietal2015} have recently studied the stability of twisted and
rotating jets caused by the velocity jump near the jet surface and have derived
the dispersion relations for incompressible plasma. In these two studies the
azimuthal flow is assumed to be invariant in the azimuthal direction and
stationary in time in order to find analytic dispersion relations. However, in
reality the transverse oscillations of a tube are much more involved.

Here, the full time-dependent problem is solved, involving the generation of
small length-scales in the inhomogeneous layer plus  the continuous energy
transference to this layer. This process is investigated by means of numerical
simulations providing unique details about the onset of the KHI for different
magnetic tube configurations. The method adopted here does not suffer from the
limitations of the models of \citet{heypri83,brownpri84} and is not restricted to
the stationary state of the system as in \citet{soleretal2010, temurietal2015}.
This allows us to perform, for the first time, a quantitative analysis of the
effect of magnetic twist on the development of the KHI in the context of
line-tied oscillating magnetic flux tubes, representing for example, coronal
loops, that undergo resonant damping while they are oscillating.

\section{Cylindrical twisted flux tube model}\label{model}

We consider a straight twisted magnetic tube of radius $R$ and length $L$ homogeneous in the axial direction. The equilibrium magnetic field depends
on $r$ only and has two components, axial, $B_z$\/, and azimuthal, $B_\phi$\/. The force-free condition in the absence of pressure gradients (we assume a constant gas pressure) reduces to
\begin{equation}
\frac{dB^2}{dr} = -\frac{2B_\phi^2}r ,
\label{forcefree}
\end{equation}
where $B^2 = B_z^2 + B_\phi^2$\/. We assume that $B_{ze}$ is
constant, $B_{\phi i}$ is proportional to $r$\/, $B_{\phi e}$ is inversely
proportional to $r$\/, and the magnetic field is continuous at $r = R$ (the subscripts {\em e} and {\em i} refer to the external and internal values, respectively). This model has been recently used by \citet{ruderterr2015,giaketal2016}. Then,
using Eq.~(\ref{forcefree}), we obtain that the equilibrium magnetic field is given
by
\begin{equation}
B_z^2 = \left\{\begin{array}{cl} B_0^2 + 2A^2(1 - r^2/R^2), & r < R, \vspace*{1mm}\\ 
   B_0^2, & r > R, \end{array}\right. 
\end{equation} 

\begin{equation}
B_\phi = \left\{\begin{array}{ll} A r/R , & r < R. \vspace*{1mm}\\ 
   A R/r, & r > R. \end{array}\right.
\label{btwist}
\end{equation}
\noindent where $A$ and $B_0$ are positive constants. In this model
the ratio of $B_\phi/B_z$ at $r=R$ is $A/B_0$. In the present work four
different values of twist are considered, $B_\phi/B_z=[0,0.1,0.2,0.3,0.4]$ (at
$r=R$) and the corresponding twist angles are $\theta=[0,5.7,11.3,16.7,21.8]$
(in degrees). The number of turns over the cylinder length $L$ is defined as
\begin{equation}
N_{tw}=\frac{1}{2 \pi} \frac{L}{r} \frac{B_\phi}{B_z}.
\label{twistntw}
\end{equation}

\begin{figure}
\plotone{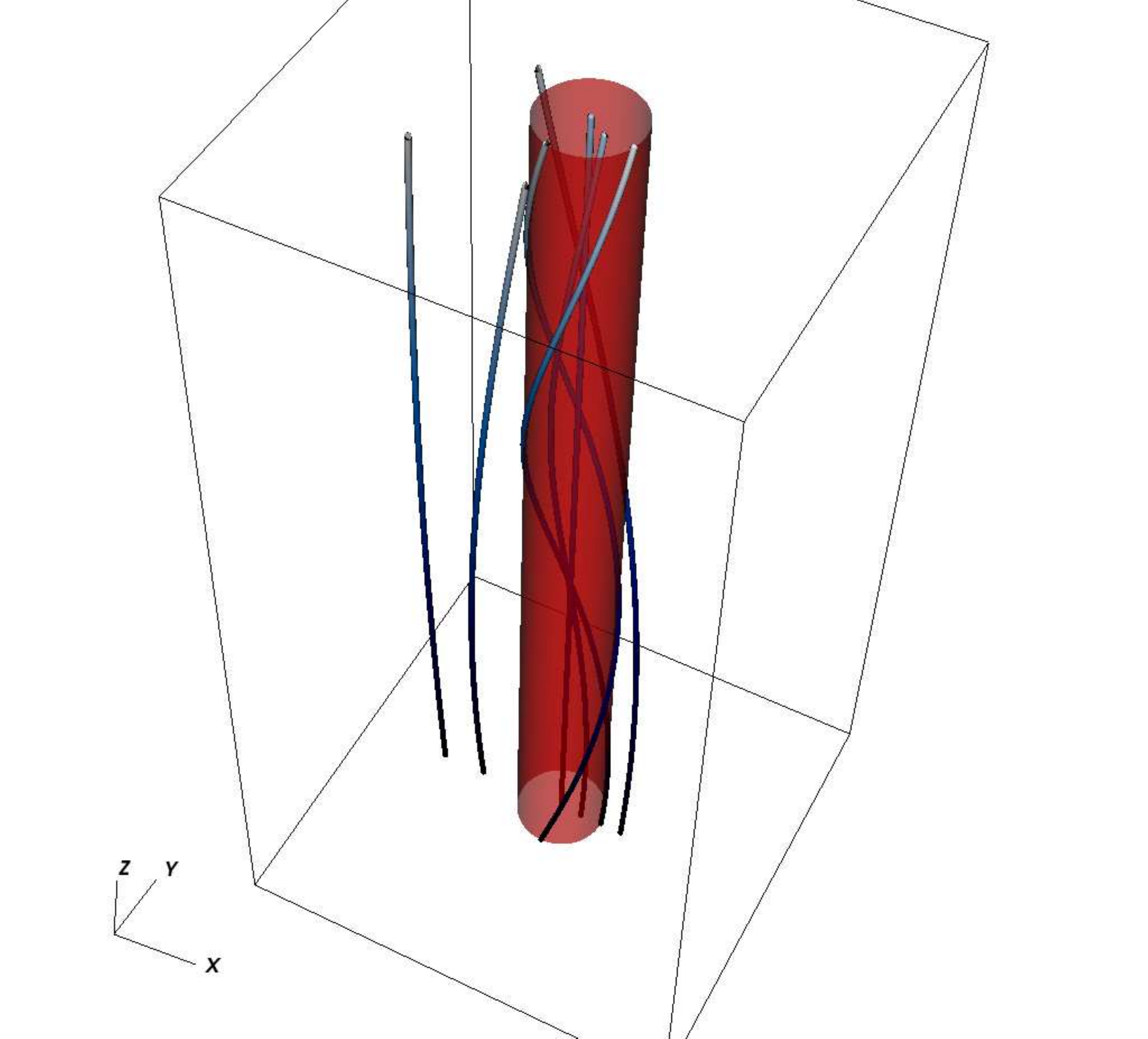}
\caption{\small Sketch of the equilibrium model. Some specific magnetic field
lines for the case $B_\phi/B_z=0.3$ are plotted in blue. The maximum number of
turns over the cylinder is 0.95. The density isocontour is plotted in red and
corresponds to $r=R$.}\label{equil}
\end{figure}

\noindent At $r=R$, where twist has a maximum, we have that
$N_{tw}=[0,0.32,0.64,0.95,1.27]$ (assuming that $L=20R$). An example of the
magnetic field configuration is shown in Fig.~\ref{equil}. We have intentionally
avoided a higher number of turns because the magnetic tube might become kink
unstable. In fact, according to the results of \citet{hoodpriest1979} for a force-free field of uniform twist the critical twist is 3.3$\pi$, which is 1.65 turns. Therefore, we are not too far from this critical value. The range of values for the twist considered in this work are in
agreement with the twist angles that are inferred from the estimations performed
by \citet{aschwanden2013,aschwandenetal2014,aschwandenetal2016}. These
estimations are based on nonlinear force-free magnetic fields calculated using
line-of-sight magnetograms and the geometry of observed coronal loops in 3D.
The twist angle can be calculated from the expression $\theta=\rm
atan\left(\sqrt{q_{free}}\right)$ where $q_{free}$ is the free magnetic energy
ratio which is typically in the range  $0<q_{free}<0.2$ \citep[see for example
table IV of][]{aschwandenetal2014}  meaning that
the twist angle is always below $24^{\circ}$.

The equilibrium plasma density, $\rho$, is assumed to be constant inside
($\rho_i$) and outside the tube ($\rho_e$) but with a smooth density transition
of thickness $l$ in the radial direction. We use a sinusoidal profile to connect
the internal and external regions of the tube \citep[see for
example,][]{rudrob02}. Other density profiles have been also considered in the
literature \citep[see][]{soleretal2013c,soleretal2014}. In this work three
different widths of the inhomogeneous layer are considered, $l/R=[0.3, 1 ,2]$,
covering the situation of an intermediate layer ($l/R=0.3$), a thick 
layer, and  a fully  inhomogeneous tube ($l/R=2$). The situation with a
discontinuous jump in density, $l/R=0$, is not considered in this work since from
the numerical point of view it is the most difficult case (the jumps in some of
the variables are artificially smoothed by numerical diffusion). A similar
problem arises if the layer is too thin, and for this reason in the present work
we have preferred to resolve well the layer and have only considered layers with
$l/R\ge 0.3$. Nevertheless, the behavior for thin layers can be inferred from our study.

\section{MHD equations and initial excitation}\label{eqsinit}

The ideal three-dimensional magnetohydrodynamic (MHD) equations are numerically solved. The equations are the following, 
\begin{eqnarray*}
\frac{\partial{\rho}}{\partial t}+\nabla\cdot \left({\rho \bf v}\right) =0,
\end{eqnarray*} \begin{eqnarray*} \frac{\partial{\rho \bf v}}{\partial
t}+\nabla\cdot \left({\rho \bf v \bf v} +P {\bf I}-\frac{\bf B \bf
B}{\mu}+\frac{{\bf B}^2}{2 \mu} \right) =0, \end{eqnarray*}
\begin{eqnarray*} \frac{\partial{\bf B}}{\partial t}=\nabla \times \left(\bf v
\times \bf B \right),   \end{eqnarray*} \begin{eqnarray*}
\frac{\partial{P}}{\partial t}+\nabla\cdot \left({\gamma P \bf v}\right) =
\left(\gamma-1\right) {\bf v}\cdot \nabla P, \end{eqnarray*}

\noindent where $\bf I$ is the unit tensor and the rest of the symbols have their
usual meaning. Temperature does not appear explicitly in the previous equations
but it is  easily calculated using the ideal gas law and the values of pressure
and density. Gas pressure, denoted by P, is included in the model mainly to
avoid significant ponderomotive forces associated to nonlinear phenomena that may
lead to the destruction of the tube because of the continuous accumulation of
density at half the loop length \citep[see][]{rankinetal94,terrofm04}. The
plasma-$\beta$ in the present model is around 0.2 and gas pressure is initially
uniform in the whole domain.

The initial excitation is specifically chosen to deposit most of the energy in the
fundamental mode along the tube axis, i.e., with longitudinal number $k_z=n\, \pi/L$ with $n=1$. For this reason,
and using the results of  \citet{ruderterr2015}, the perturbation has the following
dependence,
\begin{eqnarray}  v_x &=& f(x,y)\, \cos\left(\frac{A}{B_0 R }\frac{1}{4} \left(2
z-L\right)\right) \,\sin\left(\frac{\pi}{L} z\right),\label{twistpol1} \\  v_y
&=& f(x,y)\, \sin\left(\frac{A}{B_0 R }\frac{1}{4} \left(2 z-L\right)\right)
\,\sin\left(\frac{\pi}{L} z\right).\label{twistpol} \end{eqnarray}   

\noindent Strictly speaking there should be also a perturbation in the vertical
component of the velocity, $v_z$. However, for the values of twist considered in
this work this velocity component is rather small in comparison with $v_x$ and
$v_y$, and therefore is not included in the initial excitation.

In the absence of twist, $A=0$, the initial perturbation is in the $x-$direction only and purely
sinusoidal along the $z-$coordinate. When $A\neq 0$ the polarization is in general mixed and it
depends on the position along the tube axis, although at $z=L/2$ is still in the $x-$direction. It is
interesting to note that the kinetic energy (proportional to $v_x^2+v_y^2$) of the chosen perturbation
is independent of the amount of twist, therefore, a direct comparison between different equilibria can
be performed.

The shape of the perturbation in the plane $x-y$ is chosen to have a Gaussian
profile centered on the tube axis,  \begin{eqnarray}  f(x,y)=v_0\, 
e^{-\left(x^2+y^2\right)}, \label{shape}\end{eqnarray}
where we have taken $R=1$ for numerical convenience.  This perturbation
produces a lateral displacement of the whole loop, mostly representing the
transverse $m=1$ kink oscillation when the density has a jump at the tube
boundary. An important parameter in this study is the amplitude of the initial
velocity, $v_0$, which will determine, together with some other quantities, the
degree of nonlinearity of the physics in our problem.

\section{Numerical tools}

\subsection{MHD simulations}\label{simul}

The numerical scheme used in this work to solve the time-dependent MHD equations
is similar to that in \citet{terradasetal2016}. It is an evolution of the code
MoLMHD in which a hybrid scheme is adopted. Density, pressure and the three
components of the velocity are solved with a 6$^{th}$ order WENO scheme while
for the magnetic field a finite difference scheme is used. 

Although the initial equilibrium model is described in cylindrical coordinates,
the nonlinear MHD equations are solved in a Cartesian coordinate system. Hence,
the appropriate transformation between cylindrical and Cartesian coordinates is
used. The computational domain used in the simulations is $-5<x/R<5$, $-5<y/R<5$,
and $0<z/R<20$. Three different numerical resolutions have been considered in the
present work. Low resolution, with [200, 200, 40] points in the [$x, y, z$]
coordinates, medium, with [400, 400, 80] points, and high, with [800, 800, 80]
points. In the $z-$direction the resolution is lower than in the $x-y$ plane,
since the oscillations have a sinusoidal profile with a typical wavelength of
$2L$. However, the presence of twist introduces additional length-scales
along to the  $z-$direction, and for this reason some simulations have been
performed with an improved resolution of 320 points in the $z-$direction. 

Regarding the boundary conditions, line-tying conditions are applied at the planes $z=0$ and $z=L$
($20R$), while extrapolated conditions (zero derivative) are imposed at the rest of the boundaries
allowing energy to leave the system. Note that the tube can be considered as thin since $R\ll L$.

\subsection{Analysis of azimuthal wavenumbers}\label{aziman}

\begin{figure}[!hh] \center{\includegraphics[width=8cm]{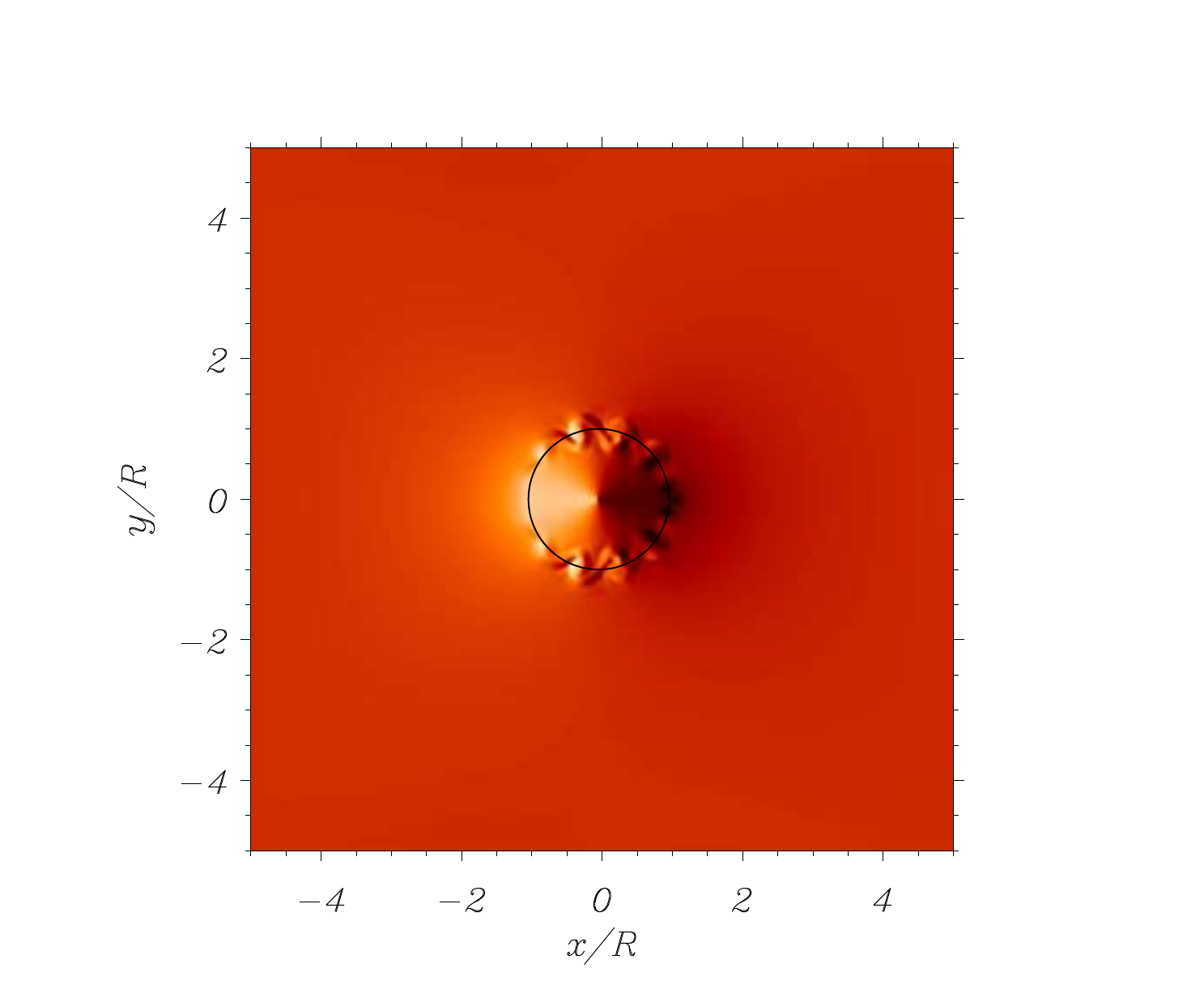}}
\caption{\small Example of the radial velocity calculated from the Cartesian
components of the  velocity using the position of the CM as the center of the
polar coordinate system. The radial velocity is interpolated on the black circle
where the Discrete Fourier Transform is computed. }\label{project} \end{figure}

One of the effects of nonlinearity in the present problem is the excitation of
different azimuthal wavenumbers. We have applied a local Fourier decomposition
to quantify the contributions of the azimuthal numbers as a function of time.
First, we have calculated the position of the center of mass (CM) of the system
(at the plane $z=L/2$, i.e., half the loop length). Since the initial
perturbation (Eqs.~(\ref{twistpol1})-(\ref{shape})), mostly excites the
transverse $m=1$ mode, the CM oscillates periodically with time. This magnitude
has been used to define the center of our local polar coordinates, meaning that
the variables of interest, in our case $v_x$ and $v_y$, have been transformed
into $v_r$ and $v_\phi$. In the linear regime the position of the tube boundary
is simply located at $r=R$ if we follow the motion of the tube axis. However, in
the nonlinear regime, wavenumbers with $p>1$ are excited, leading to changes in
the shape of the tube boundary (we use $p$ instead of $m$ to distinguish between the
wavenumbers excited in the simulations and the eigenmode solutions). We have
calculated the profile of $v_r$ and $v_\phi$ along the circular path at $r=R$ 
($v_{r R}$ and $v_{\phi R}$) in the moving polar coordinate system, i.e.,
following the motion of the CM. An example is shown in Fig.~\ref{project}.

Then we have used the Discrete Fourier Transform to analyze the contribution of
the different values of $p$ using the estimated profiles of $v_{r R}$ and
$v_{\phi R}$ (namely the function $g$), i.e., \begin{eqnarray} 
G(p)=\frac{1}{N}\sum_{k=0}^{N-1}g(k)\,e^{-i\frac{2\pi}{N}p\, k},\label{fourier}
\end{eqnarray} for a discrete set of $N$ samples ($p=0,\ldots,  N-1$). In our case,
the analysis is in the azimuthal direction, ranging from 0 to $2\pi$, meaning
that instead of $k$ it is more convenient to introduce the parameter
$\theta_k=2\pi k/N$. Therefore, with the aid of Eq.~(\ref{fourier}) we know the
contribution of each $p$ to the total signal, which is, using the inverse
Fourier transform, \begin{eqnarray}  g(\theta_k)=\sum_{p=0}^{N-1}G(p)\, e^{i\,
p\, \theta_k }.\label{fourierinv} \end{eqnarray} Since the spectrum is
symmetric about $N/2$ we only need to consider $p=0,\ldots, N/2$ in our analysis.
Due to the parity of $v_{r R}$ and $v_{\phi R}$ the corresponding $G(p)$ is
purely real and purely imaginary, respectively. In the example of
Fig.~\ref{project},  $p=1$ is dominant but the excitation of high order $p$s is
evident.

\section{Results for the untwisted tube}\label{notwist}

We start with the magnetic configuration without twist. This will help to
understand and interpret the results of the twisted magnetic tube,
presented in the following sections. 

In the simulations, we excite the tube by imposing the initial velocity profile
given in  Section~\ref{eqsinit}. It is useful to know the corresponding
spatial displacement in order to assess the degree of nonlinearity expected in
the system.  In the linear regime the relation between the tube velocity and
the tube displacement is $v_0=\omega\,\xi_0$, where $\omega$ is the frequency
of oscillation. For the transverse kink mode in the limit of $R/L\ll 1$, i.e.,
the thin tube approximation, we have that $\omega_{\rm k}=k_z\, c_{\rm k}$ where
$k_z=\pi/L$ for the fundamental mode and $c_{\rm k}$ is the kink speed. In terms
of the internal Alfv\'en velocity we have that \begin{eqnarray}\label{ckink}
c_{\rm k}=\sqrt{\frac{2}{1+\rho_e/\rho_i}}\, v_{Ai}.  \end{eqnarray}  The relation
between velocity and displacement is therefore \begin{eqnarray}\label{vlin}
\frac{v_0}{v_{Ai}}=\pi \sqrt{\frac{2}{1+\rho_e/\rho_i}}\, \frac{\xi_0}{L}.
\label{voamp} \end{eqnarray} For the transverse kink mode $v_0$ and $\xi_0$
can be viewed as the maximum velocity and maximum displacement of the tube axis
with respect to the equilibrium position. As explained in \citet{rudgos2014}
the parameter associated to the nonlinearity of the problem is indeed
$\xi_0/{R}$ and not $\xi_0/{L}$. Therefore, the  linear theory is valid when
$\xi_0/{R} \ll 1$, i.e., when the maximum displacement of the tube is small in
comparison with the radius of the tube.

Using Eq.~(\ref{voamp}) we can write   \begin{eqnarray}\label{xilin}
\frac{\xi_0}{R}=\frac{L}{R} \frac{v_0}{v_{Ai}}\frac{1}{\pi}
\sqrt{\frac{1+\rho_e/\rho_i}{2}},\label{xioamp}   \end{eqnarray} which shows
that if $v_0/v_{Ai}\ll 1$ does not necessarily mean that $\xi_0/{R}\ll 1$. This
magnitude  depends also on the ratio $L/R$ which is in general quite large, at
least for the observed oscillating loops in the solar corona. Hereafter, we
perform a parametric study changing the main parameter of our problem,
$\xi_0/{R}$. 

\subsection{Linear results}

\begin{figure}[!hh] \center{\includegraphics[width=7cm]{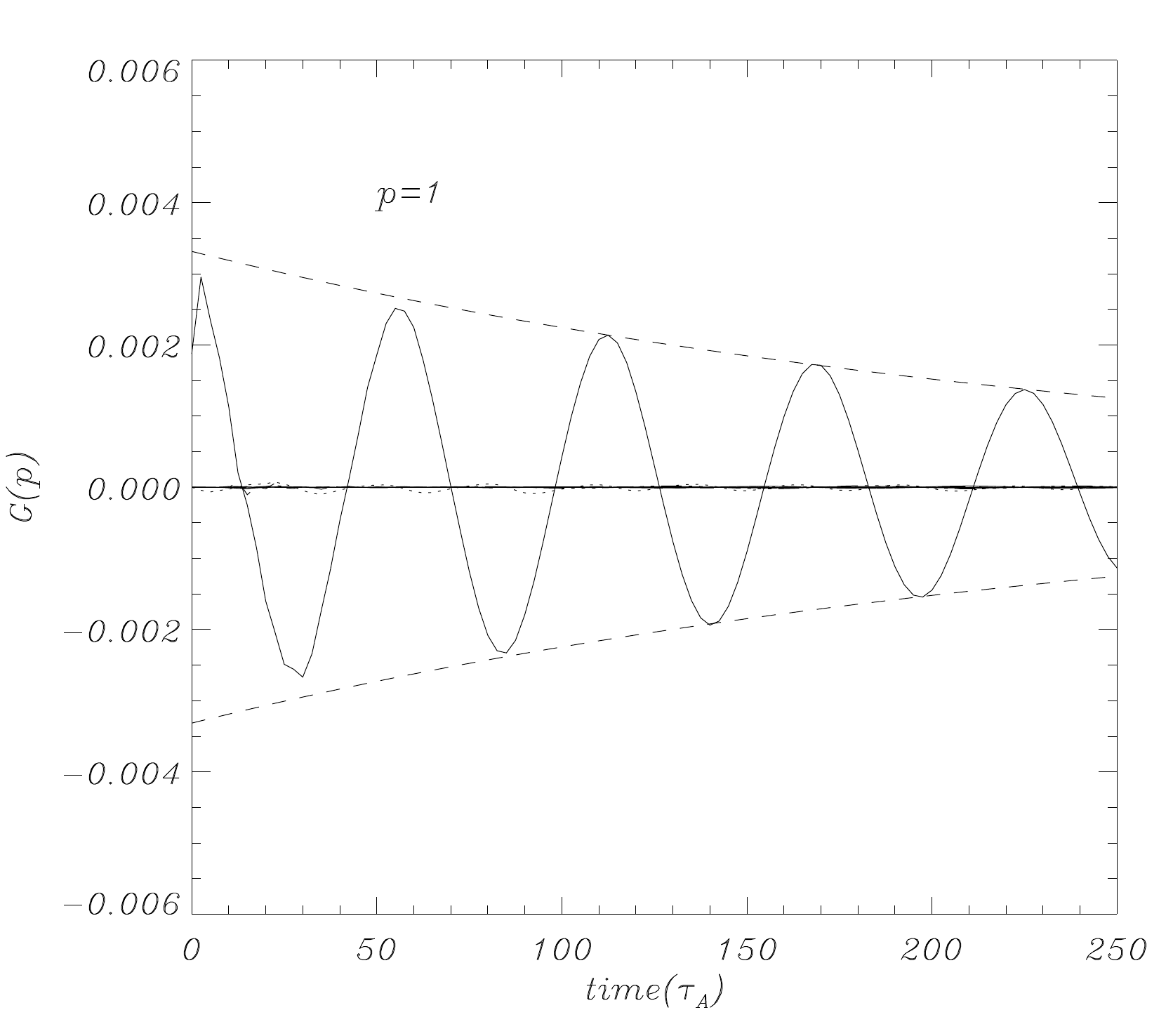}}
\center{\includegraphics[width=7cm]{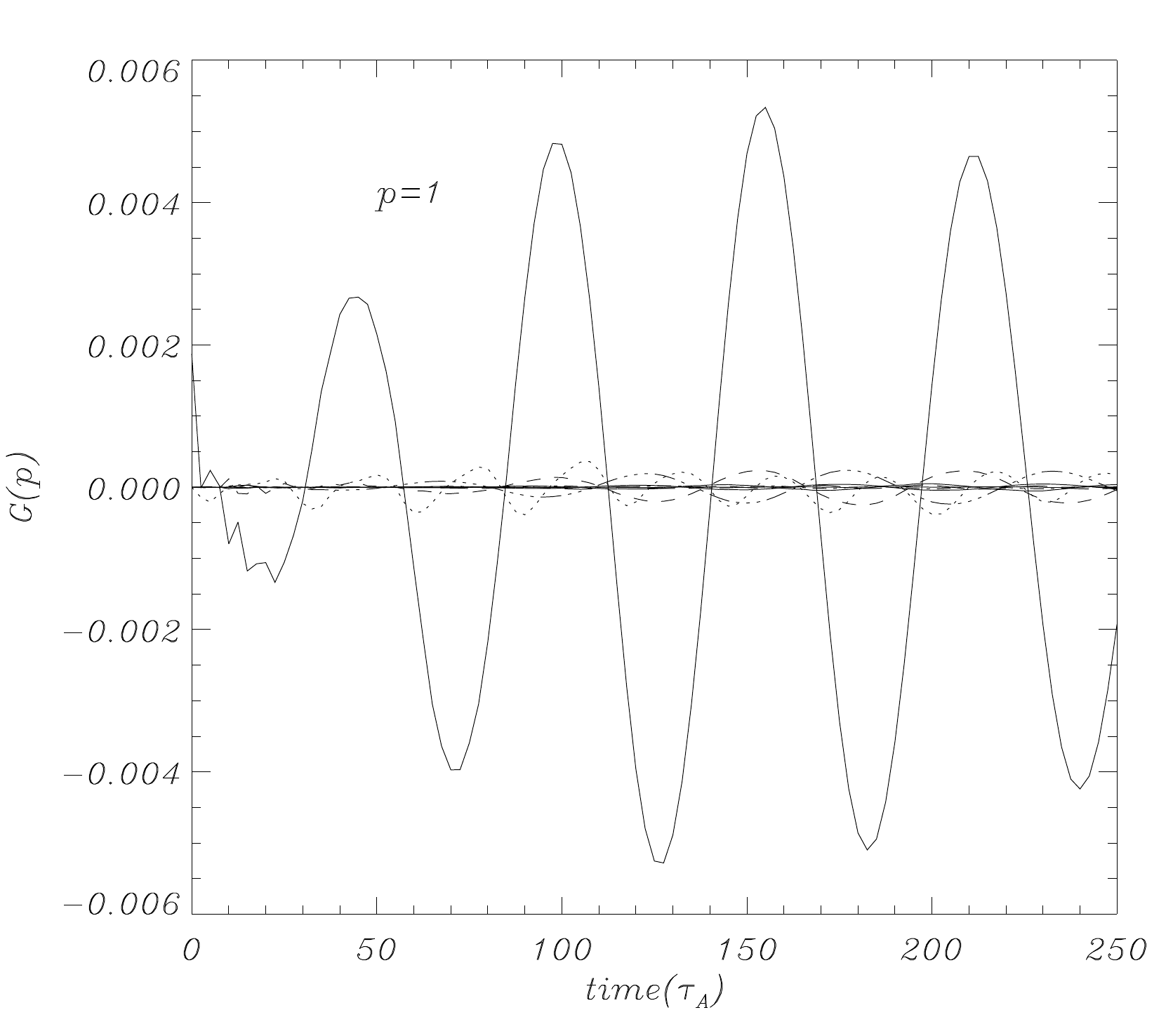}} \caption{\small
Fourier coefficients as a function of time for the radial velocity $v_{r
R}$ (top panel) and azimuthal velocity $v_{\phi R}$ (bottom panel). The
contribution of the $p=1$ is dominant. Higher values of $p$ (plotted with
dotted and dashed lines) are due to weakly nonlinear effects and are more
clear in the bottom panel. In the top panel, an exponential fit using the
last three peaks has been performed to estimate the damping time. In this
case $l/R=0.3$ and there is no twist.}\label{Fouriercoeflin} \end{figure}

As a method of testing the simulations and numerical method, we start our study with the analysis of
the evolution of the system for a perturbation in the linear regime, $\xi_0/{R}=0.05\ll 1$. We perform
the analysis of the azimuthal wavenumbers explained in Section~\ref{aziman} on the output of the
numerical simulations. The results are plotted in Fig.~\ref{Fouriercoeflin}. We obtain the expected
behavior, the tube is oscillating laterally around the equilibrium position at the kink frequency
($\omega_{\rm k}$). The sinusoidal motion of the tube axis is slightly attenuated with time (see
dashed line in top panel) and it is due to the process of continuum damping associated to the presence
of the inhomogeneous layer between the core of the tube and the external medium
\citep[see the time-dependent analysis in][]{terretal06b}. At the same time the amplitude of the azimuthal velocity component grows
with time (see bottom panel) due to the energy that is pumped into the layer. The increase in the
amplitude of the azimuthal velocity component is bounded since the initial perturbation has a finite
energy. The attenuation at the end of the time interval is due to numerical dissipation. 

The agreement between the results of the simulation and the theoretical predictions is remarkable. The
exponential decay given by the analytical expressions for the damping rate work very well after two or
three periods while at the beginning of the evolution it is better fitted by a Gaussian profile
\citep[][]{pascoetal2012,hoodetal2013,rudermanterradas2013}. From the simulations we infer a damping
per period of 4.3 while the analytical theory provides a value of 4.1 \citep[see the analytic
expressions for the thin tube (TT) and thin boundary (TB) approximations in for example,
][]{rudrob02}. The good match we obtain in our runs is similar to that of  \citet{magyartom2016} in
the linear regime. From Fig.~\ref{Fouriercoeflin} bottom panel we appreciate a very weak excitation of
higher $p$s which is related to nonlinear effects described in detail in the following sections.

\subsubsection{Effect of numerical resolution}

An increase in the grid resolution mostly affects the small scales that are
generated in the inhomogeneous layer while the attenuation times of the
oscillations are essentially unaltered. This has been checked comparing the
results for medium and high resolutions. This means that the  dissipation scales
are much shorter than the global characteristic scale of the transverse kink
motion. This behavior agrees with the results of the quasi-mode calculations that
indicate that for small dissipation coefficients (resistivity or viscosity) the 
damping times are independent of the amount of dissipation in the system
\citep[e.g.,][]{poedtskerner91,vandetal04,terretal06b,soleretal2013c}. An
increase in the resolution reduces the numerical dissipation in the azimuthal
velocity component found at the end of the time series in
Fig.~\ref{Fouriercoeflin} bottom panel.

\begin{figure}[!hh] \center{\includegraphics[width=7cm]{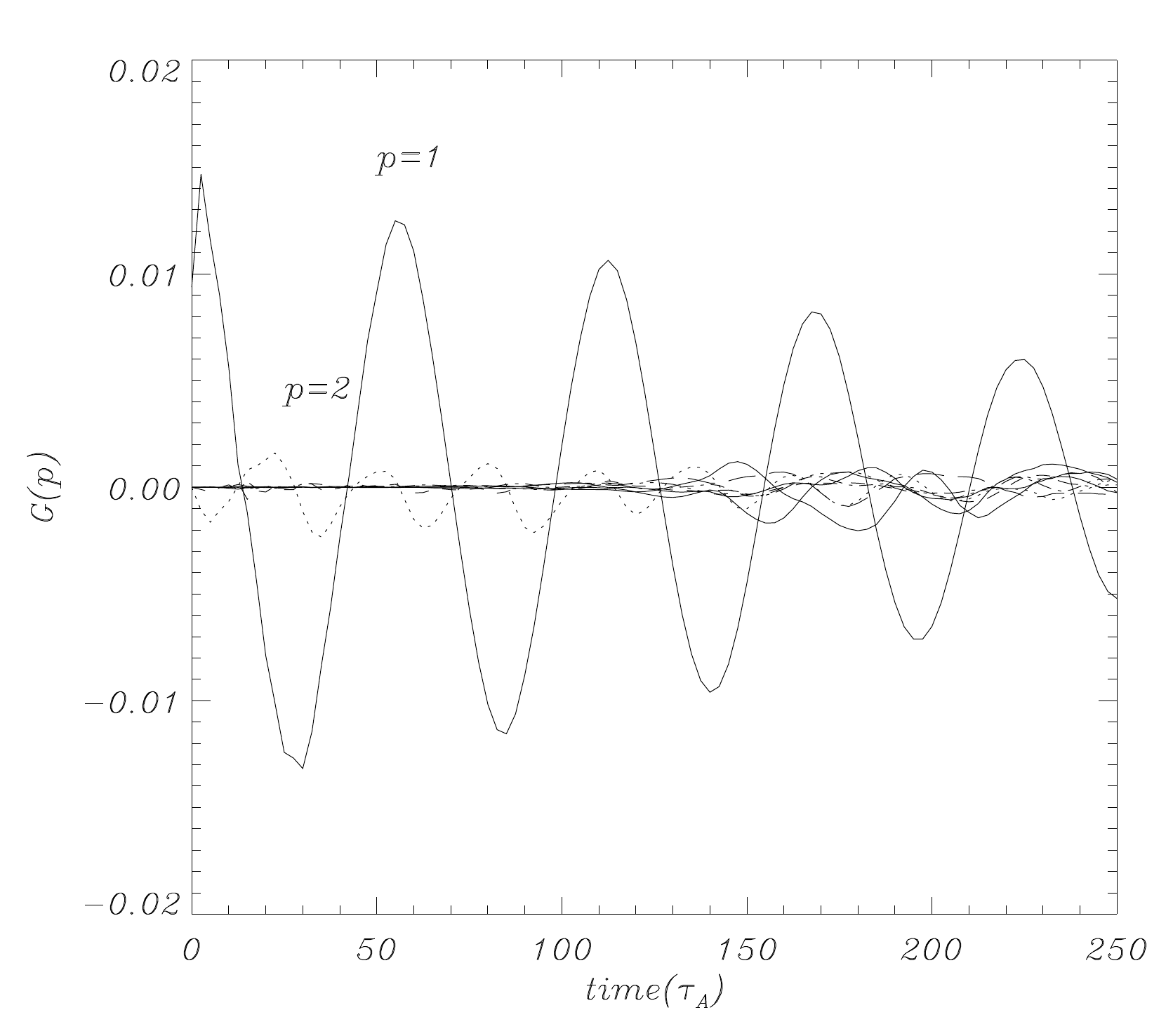}}
\center{\includegraphics[width=7cm]{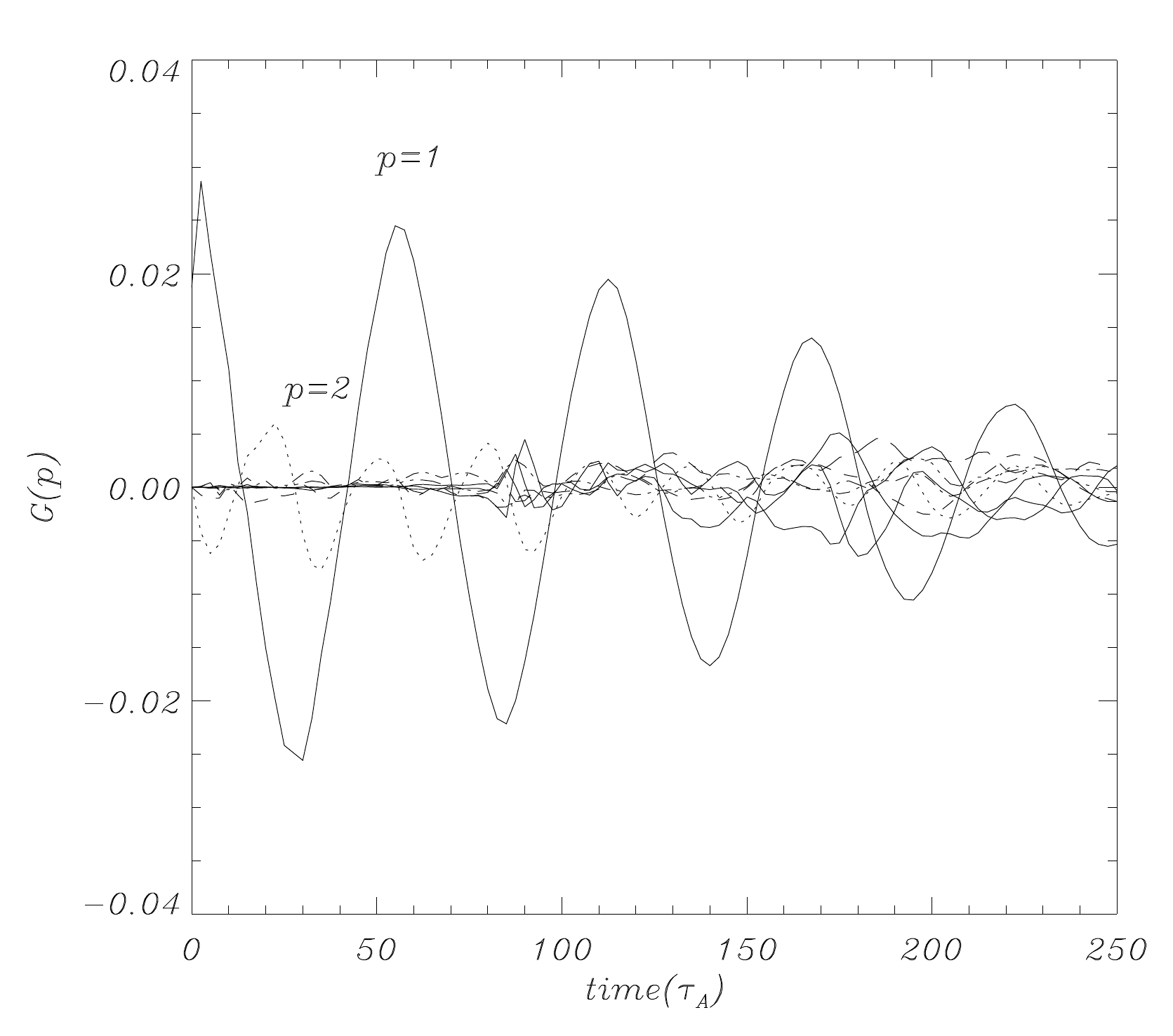}} \caption{\small Fourier
coefficients as a function of time for the radial velocity $v_{r R}$. The
contribution of the $p=1$ is dominant. Higher values of $p$ are due to
nonlinear effects. The $p=2$ is oscillating at half the period of $p=1$. In
this case $l/R=0.3$ and there is no twist. In the top panel the amplitude
of the initial excitation ($\xi_0/R=0.2$) is half the amplitude in the
bottom panel ($\xi_0/R=0.4$). Note the different ranges in the vertical
axis of the plots.}\label{Fouriercoef} \end{figure}

\subsection{Nonlinear results}\label{nonlinnotwist}

Now we concentrate in a weakly nonlinear situation ($\xi_0/R=0.2$). It was shown
by  \citet{terradasetal08}, \citet{antolinetal2014,antolinetal2015,
magyaretal2015,magyartom2016,antolinetal2017}, that under such conditions the
tube boundary develops deformations and changes associated to the KHI. For a
proper quantification of the deformation of the tube boundary,  we perform again
the analysis of the azimuthal wavenumbers. The time evolution of the Fourier
coefficients is plotted in Fig.~\ref{Fouriercoef} (top panel). The $p=1$ has the
largest amplitude since the initial perturbation is chosen to precisely excite
this mode and it shows again attenuation with time. But now  $p=2$ has a
significant amplitude and it is excited from the beginning of the time evolution.
The excitation of higher order $p$s is also present in Fig.~\ref{Fouriercoef} top
panel but only for times larger than $t/\tau_{\rm A}=150$.

\begin{figure}[!hh] \center{\includegraphics[width=7cm]{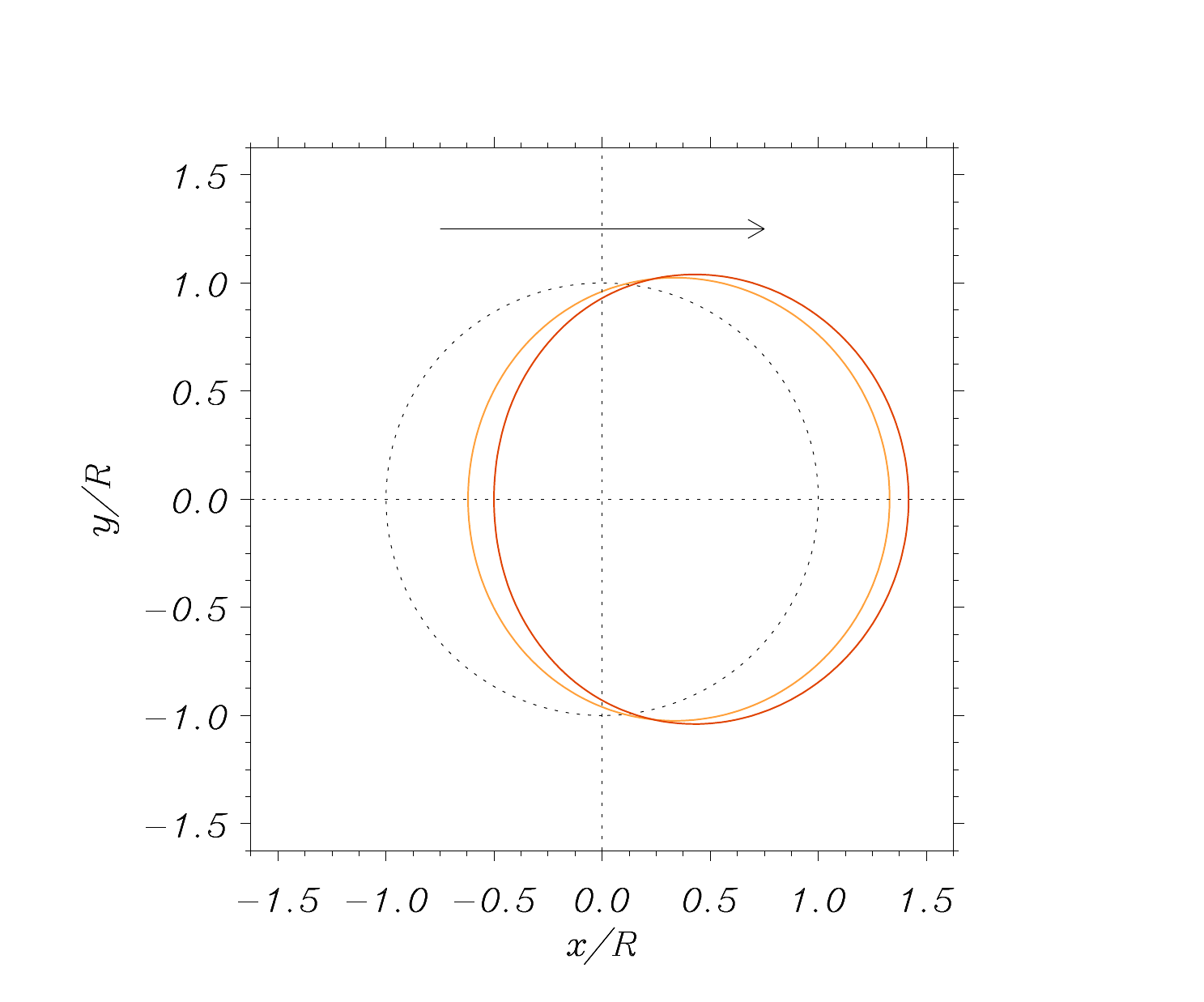}}
\center{\includegraphics[width=7cm]{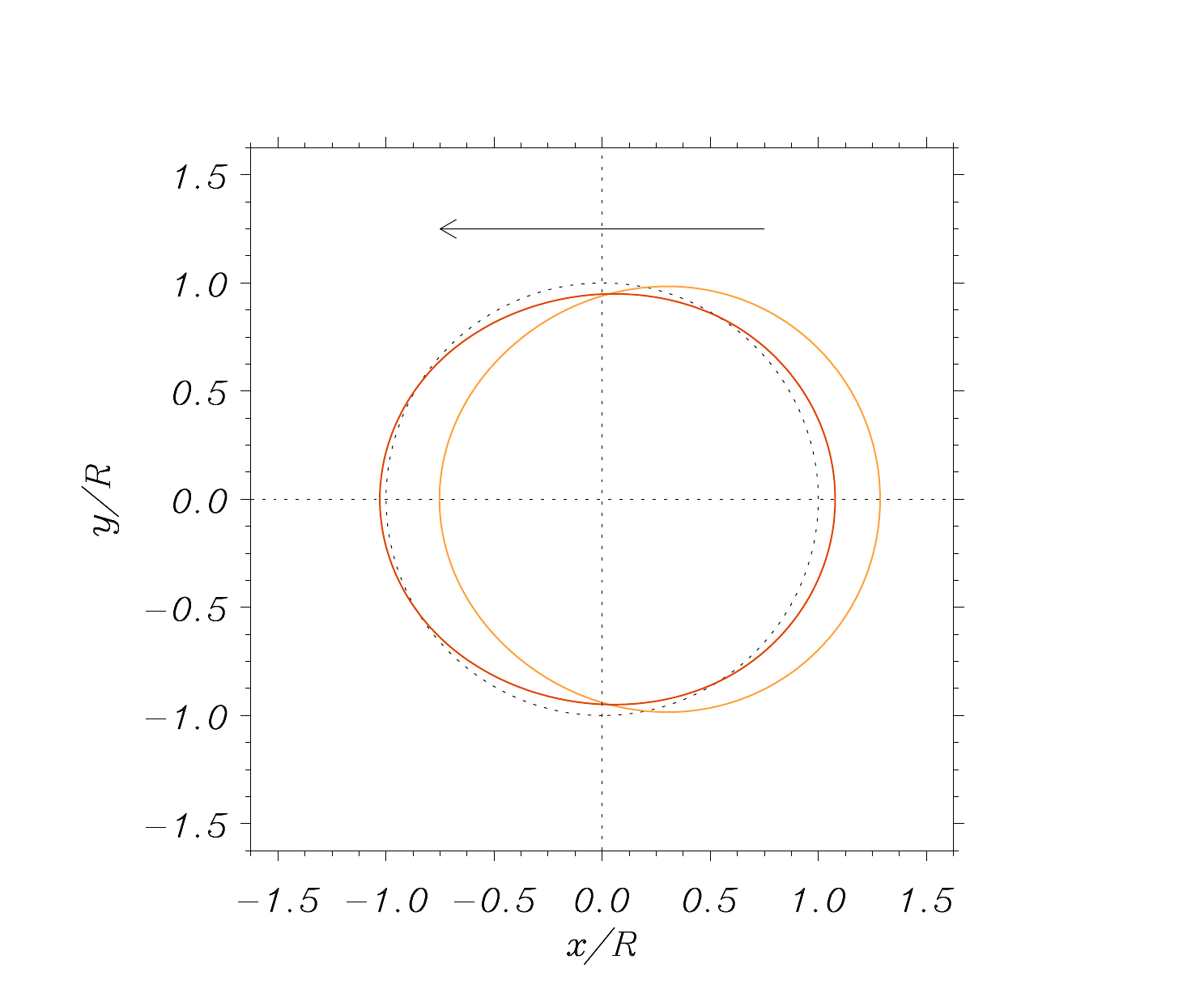}} \caption{\small Shape of
the tube cross-section at different times. In the top panel, the tube is moving in
the positive $x-$direction. In the bottom panel the motion is in the negative
$x-$direction. The excitation of the $p=2$ mode, driven by the $p=1$ mode is
clear in this plot. The circle plotted with dots represents the shape of the
tube at $t=0$.}\label{contourtube} \end{figure}

The $p=2$ has a period which is essentially half of that of $p=1$, as it can be
appreciated in Fig.~\ref{Fouriercoef}. This nicely agrees with the analytical
results of  \citet{rudgos2014}, predicting a nonlinear coupling between $p=1$ and
$p=2$ \citep[see also][for the case of propagating waves instead of
standing]{rudermanetal2010}. These authors found that the frequency (period) of
the mode $p=2$ is twice (half) the frequency (period) of the $p=1$ who acts as a
driver of $p=2$ (see their Eqs.~(82) and (83)) \citep[see also the recently
accepted paper of][]{ruderman2017}, which is confirmed in our simulation. {\bf
\citet[][]{antolinetal2017} have also described this effect in their numerical
analysis of transverse kink oscillations}. From the physical point of view, the
deformation of the edge of the tube associated to $p=2$ can be understood if we
concentrate on the plane at half the tube length at the early stages of the time
evolution, see Fig.~\ref{contourtube}. At $t=0$ the tube with a cylindrical shape
is located at $x=0$, the perturbation produces the lateral displacement of the
whole tube which immediately starts to decelerate because of the effect of the
line-tied magnetic field lines ($p=1$). The front of the tube decelerates
slightly faster than the rear since at the front the magnetic lines are more
curved, the magnetic tension is stronger, and this produces a decrease of the
width of the tube along the $x-$direction. At same time the width along the
$y-$direction increases because the motions are essentially incompressible,
meaning that the area of the cross-section has to remain constant. Once the tube
has reached its maximum displacement, the process is reverted. The width of the
tube increases in the $x-$direction and reaching the maximum elongation at the
initial position of the equilibrium.  The motion of the tube continues in the
negative $x-$direction, the width of the tube decreases in the horizontal
direction and increases in the vertical direction, recovering the situation found
at the maximum positive displacement but now the tube is located at the maximum
displacement in the negative $x-$direction. For this reason, the period of the
deformation associated to the $p=2$ has half the period of the transverse $p=1$
displacement. These explanations are based on the fact that changes in the
vertical direction of the velocity ($z-$component), perpendicular to the plane of
reference, produce a  negligible contribution to the total divergence of the
velocity, meaning that if this magnitude is basically zero then
contractions/expansions in the $x-$direction have to be balanced by equal
expansions/contractions in the $y-$direction.

\begin{figure}[!hh] \center{\includegraphics[width=7cm]{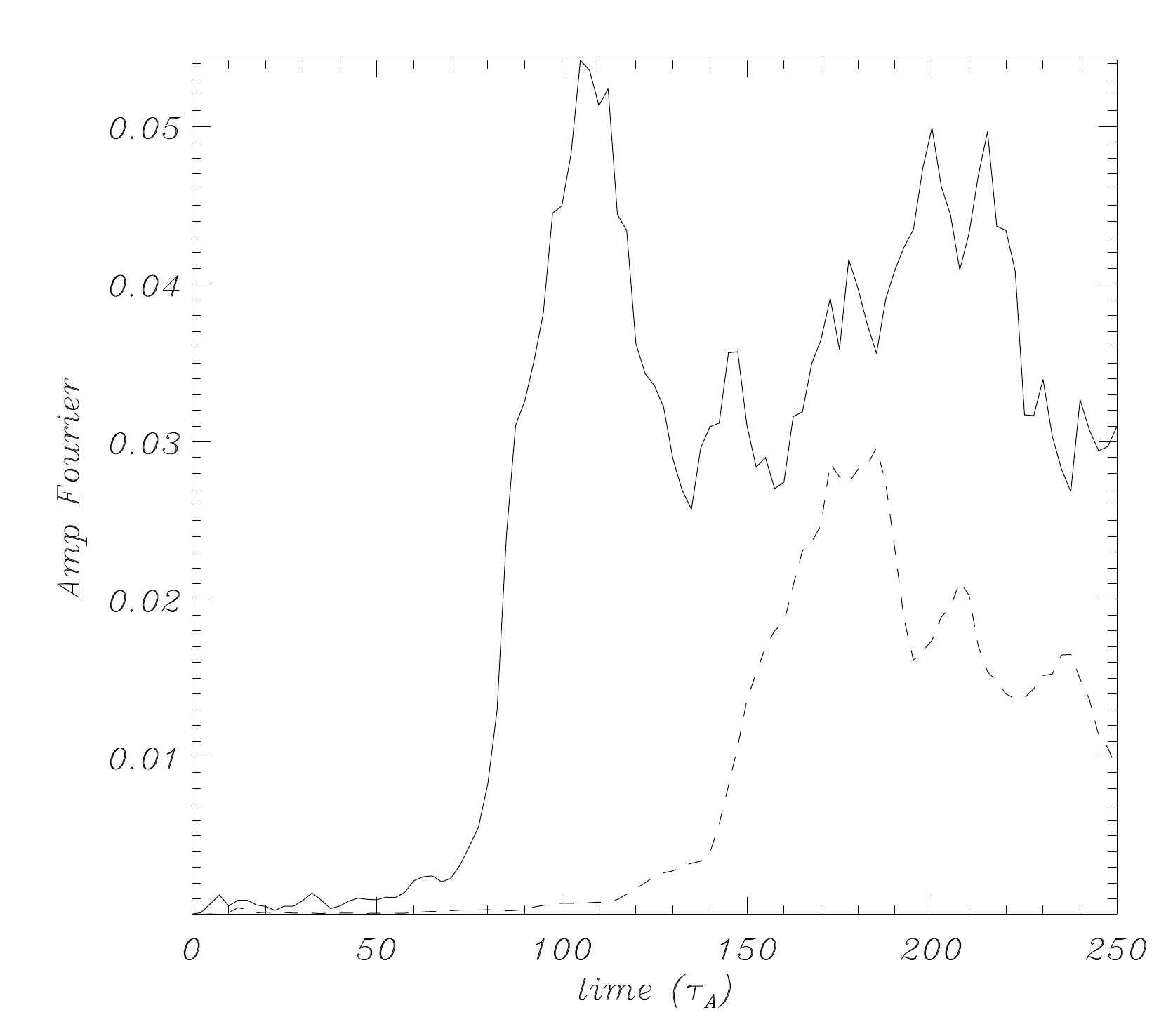}} 
\caption{\small Sum of Fourier coefficients, $\sum_{p=3}^{N/2}\left|G(p)\right|$,
as a function of time for the radial velocity $v_{r R}$. The dashed line
corresponds to an initial amplitude of $\xi_0/R=0.2$, which is half of that of the
continuous line, $\xi_0/R=0.4$ . In this case $l/R=0.3$ and there is no
twist.}\label{Fouriersum} \end{figure}

\begin{figure}[!hh] \center{\includegraphics[width=7cm]{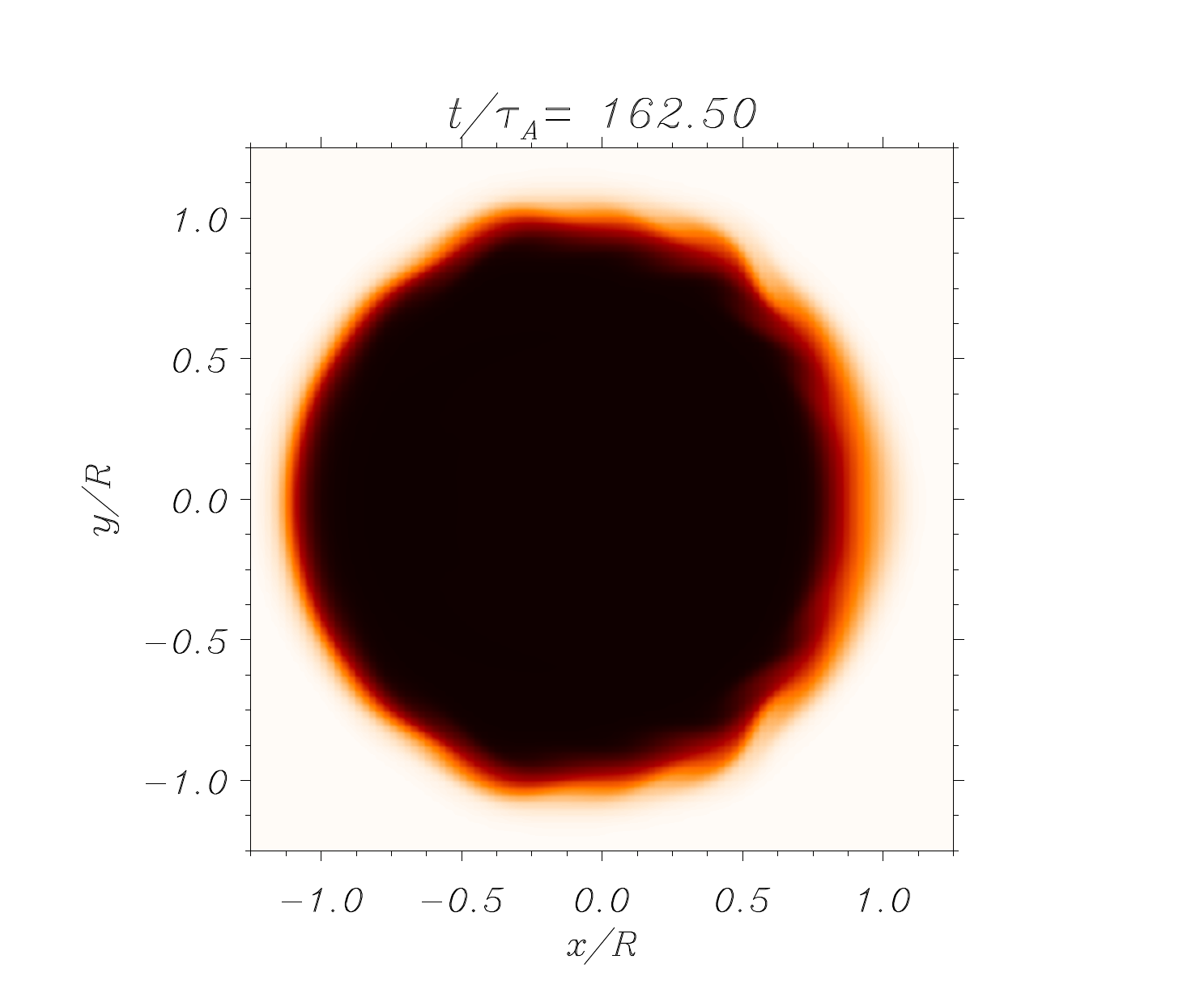}}
\center{\includegraphics[width=7cm]{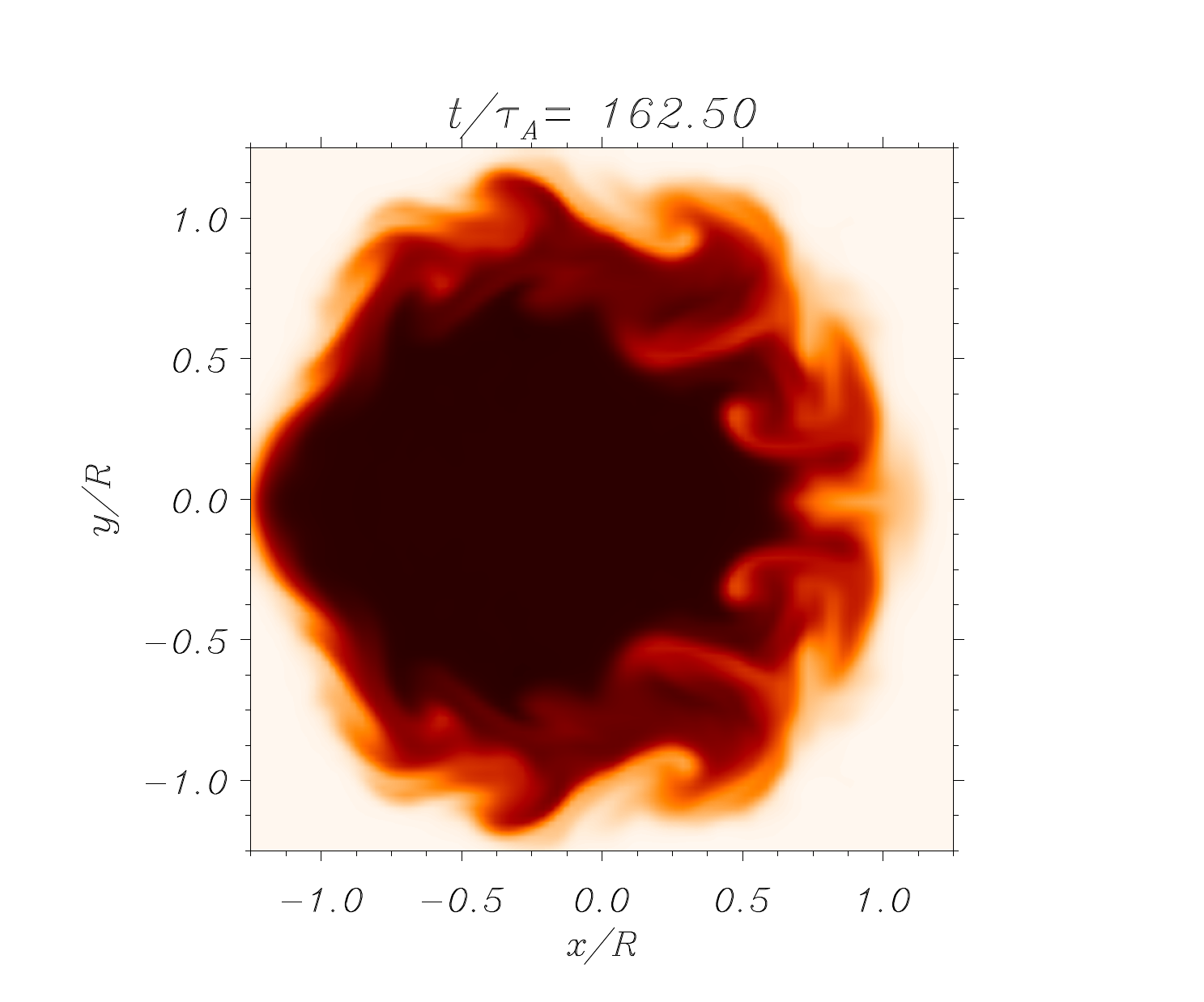}} \caption{\small Snapshot of
the two-dimensional density distribution at half the tube length ($z=L/2$)
at a given time. In the top panel the amplitude of the initial
excitation ($\xi_0/R=0.2$) is half the amplitude in the bottom
panel ($\xi_0/R=0.4$).}\label{densnotwist} \end{figure}

Interestingly, the excitation of higher order $p$s ($p$ greater than 2) in
Fig.~\ref{Fouriercoef} top panel is evident for $t/\tau_{\rm A}>150$. Contrary
to the damped behavior  of the modes $p=1$ and $p=2$ their amplitude in general
increases with time. In order to quantify this behavior we have performed a sum
of Fourier coefficients. In particular we have calculated
$\sum_{p=3}^{N/2}\left|G(p)\right|$ which is represented in
Fig.~\ref{Fouriersum} (see dashed line) as a function of time. The increase with
time of this magnitude is evident up to times around $t/\tau_{\rm A}=200$. This
behavior is due to the development of the KHI at the shear layer. The density
distribution at a given time instant is represented in the top panel of
Fig.~\ref{densnotwist}. In this plot the still weak deformations of the tube
boundary are due to the early stages of the development of the KHI.

We have repeated the run but with an initial amplitude that is twice the one in
the previous simulation. The results for $\xi_0/R=0.4$ are represented in
Fig.~\ref{Fouriercoef} bottom panel. Now the amplitude of the $p=2$ mode respect
to the $p=1$ is larger in comparison with Fig.~\ref{Fouriercoef} (top panel). In
fact, the theoretical prediction of \citet{rudgos2014} about the quadratic
dependence between the amplitude of $p=1$ and $p=2$ is also found in the
simulations, and this can be inferred from the comparison of the top panel and
bottom panel of Fig.~\ref{Fouriercoef} (the ratio of the amplitudes $p=1$ to
$p=2$ in the top panel with respect to the bottom panel, where the initial
amplitude has been doubled, is about 4). The increase of amplitude of the
nonlinear excited high order $p$s is also evident in this plot. The evolution
of the sum of Fourier coefficients is plotted in Fig.~\ref{Fouriersum}
(continuous line). The onset of the KHI is produced at earlier stages (around
$t/\tau_{\rm A}=80$) and the rise of the instability is faster. The  amplitude 
represented in  Fig.~\ref{Fouriersum} for the nonlinear case is larger than that
for the weakly nonlinear situation, meaning that more energy is eventually
deposited in high order $p$s than in the first case. The density distribution at a
given time instant is represented in the bottom panel of Fig.~\ref{densnotwist}
where significant deformations of the boundary are visible due to the excitation
of $p>2$ (compare with top panel).

\begin{figure}[!hh] \center{\includegraphics[width=7cm]{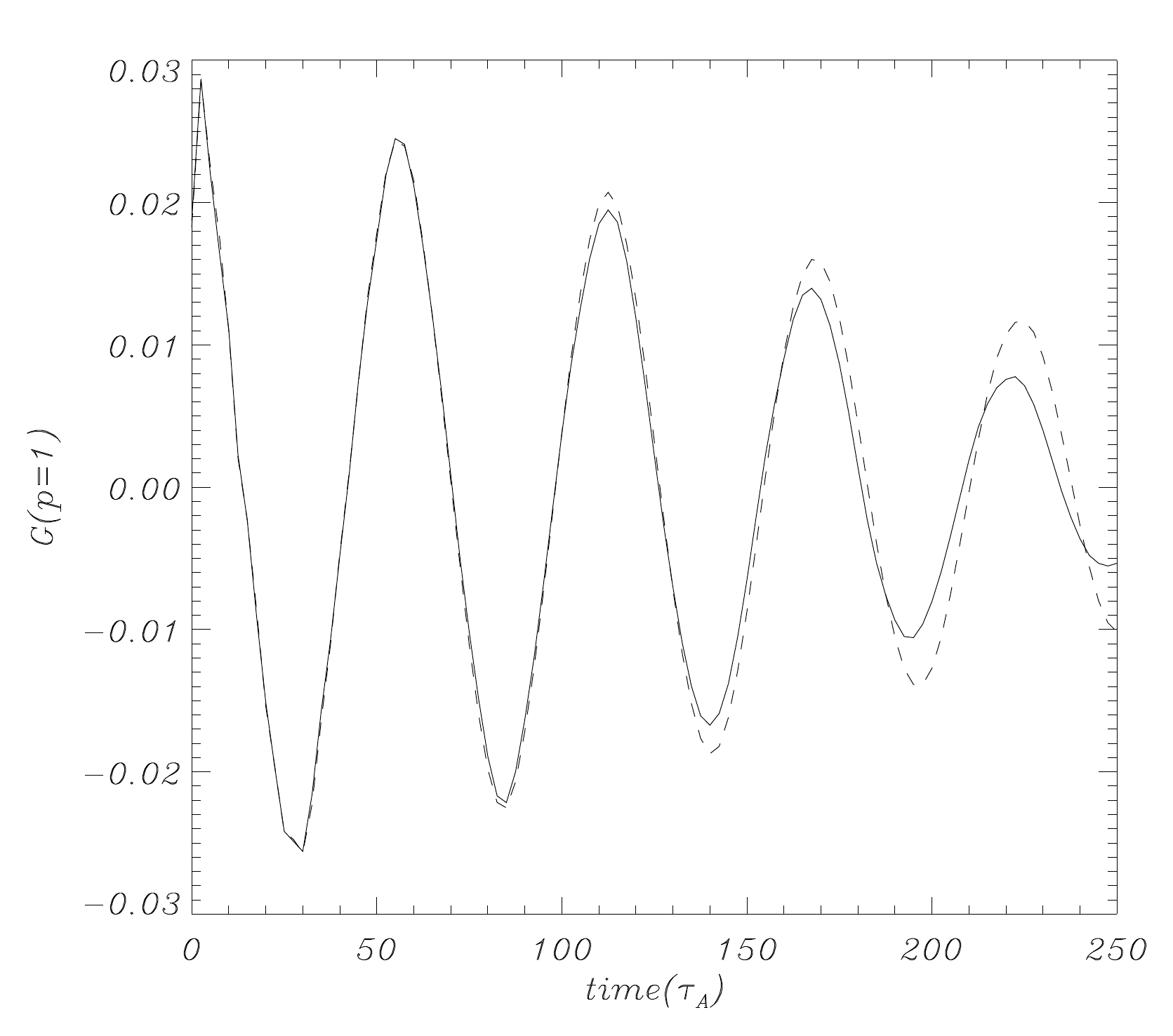}}
\caption{\small Fourier coefficients as a function of time for the radial
velocity $v_{r R}$ ($p=1$) corresponding to two different amplitudes of
excitation (same as in Fig.~\ref{Fouriercoef}). The dashed line corresponds to an
amplitude that is half of the black line, and has been normalized for comparison
purposes. The attenuation is faster when the amplitude is
larger.}\label{compardamp} \end{figure}

An additional conclusion from the comparison of the results in  Fig.~\ref{Fouriercoef} is that the
attenuation with time of the mode $p=1$ is faster for the largest initial amplitude of excitation. In
Fig.~\ref{compardamp} the two attenuation profiles for $p=1$ have been superimposed by choosing a
proper normalization factor. The plot indicates that, for times shorter than $t/\tau_{\rm A}=80$, the
two curves are almost identical. Meaning that, although the coupling with the $p=2$ is stronger for
the largest amplitude of excitation, this does not significantly affect the damping. Nevertheless, the
differences in the two curves start to be clear when the KHI develops, i.e, for   $t/\tau_{\rm A}>80$.
Therefore, the development of the KHI instability enhances the attenuation of the oscillation since
part of the energy is transferred from the initially organized flow of the mode $p=1$ to higher order
$p$s. This is in agreement with the results of \citet{magyartom2016}.

\subsubsection{Analytical estimation of KHI onset times}\label{ankhi}

The onset time for the instability in the untwisted case can be inferred
by applying the results obtained by \citet{brownpri84} in a strong phase-mixing
situation. Some caution is needed here since strictly speaking the results of \citet{brownpri84} are applicable for $m=0$ only. 

\begin{figure}[!hh]  \center{\includegraphics[width=7cm]{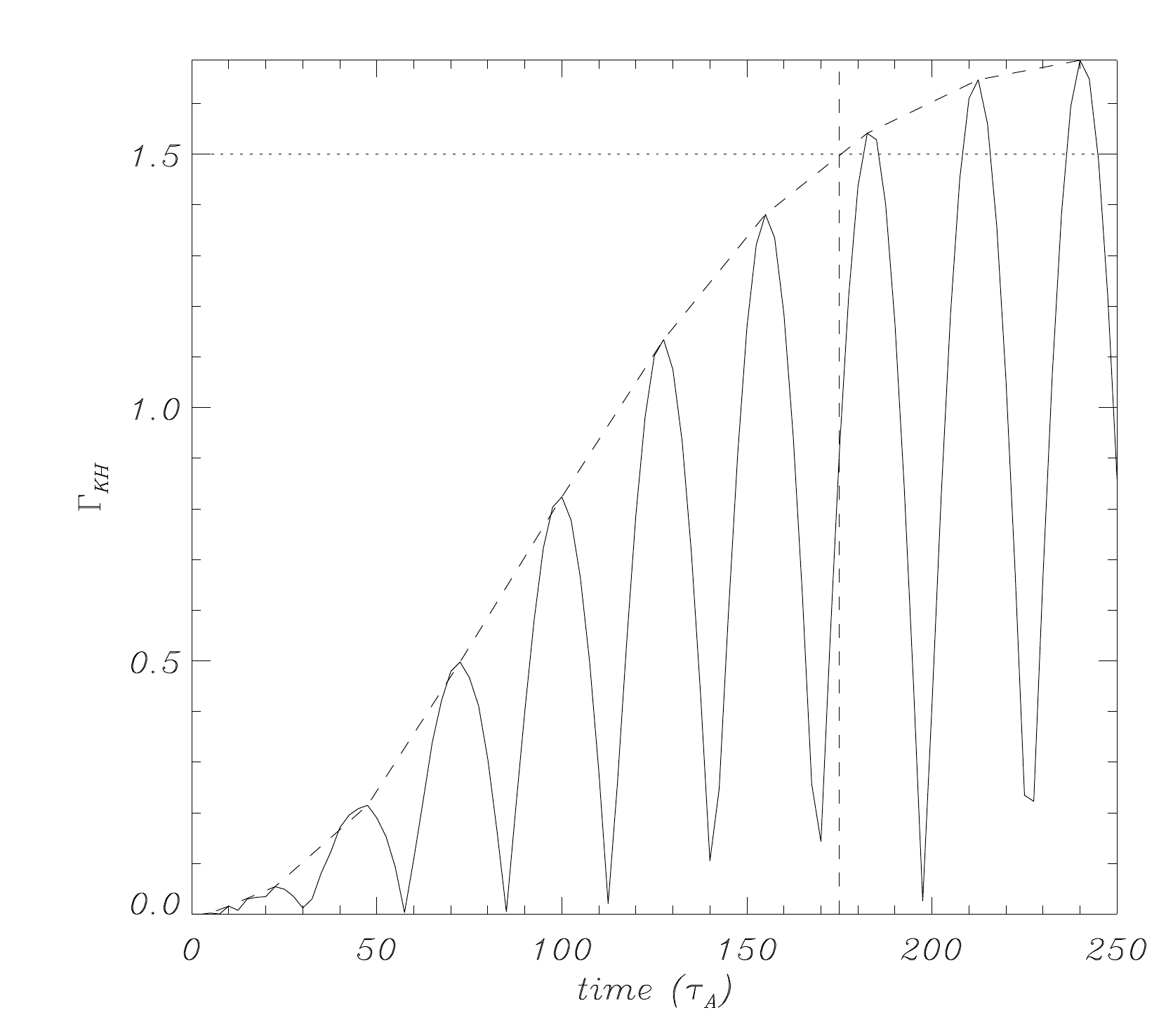}}
\caption{\small Normalized maximum growth-rate as a function of time for 
$\xi_0/R=0.2$ based on the linear results and
Eq.~(\ref{eqgammakh}).}\label{gammakhi} \end{figure}

\citet{allanwright1997} defined the following normalized maximum growth rate
based on \citet{brownpri84}, \begin{eqnarray}  \Gamma_{\rm KH}\approx
\frac{1.7\, V_0}{4\, \omega_{\rm A}\, L_{\rm ph}}, \label{eqgammakh}
\end{eqnarray} \noindent where $V_0$ is the wave amplitude, $\omega_{\rm A}$ the
frequency  at the resonance, and $L_{\rm ph}$ the phase-mixing length. The
phase-mixing length is defined as \citep[see][]{mannetal1995,wrightrick1995}
\begin{eqnarray} L_{\rm ph}=\frac{2 \pi}{\frac{d\omega_{\rm A}}{dr} t}.
\label{phasemixinglengh} \end{eqnarray}  \noindent In our model, and in the
linear regime without twist, the dependence with position of the Alfv\'en
frequency is known since we know that the density profile is sinusoidal and the
magnetic field is constant, calculating the derivative of the Alfv\'en frequency
we find that $L_{\rm ph}\approx l /(\Omega \,t)$ where $l$ is the width of the
inhomogeneous layer and $\Omega$ is a characteristic frequency. This magnitude
is time-dependent, decreasing as $t^{-1}$, because the phase-mixing process
continuously generates smaller length-scales with time. The amplitude $V_0$ at
the resonant layer depends also on time since energy is gradually pumped into
the layer until all the energy of the initial perturbation is deposited there. 
This means that $\Gamma_{\rm KH}$ is also a function of time. 
\citet{allanwright1997} defined the normalized growth-rate of the instability in
such a way that the growth of the KHI during a quarter cycle of a large velocity
shear will be significant if $\Gamma_{\rm KH}$ is comparable to or greater than
unity. Using the results of the linear simulations we compute $\Gamma_{\rm KH}$,
which is linearly proportional to the amplitude of the velocity shear at the
resonant position. If nonlinearity does not significantly modify the rate at
which the energy is pumped into the layer then we can use the profile of
$\Gamma_{\rm KH}$ in the linear regime, namely for very small amplitudes of
oscillation, to infer the onset time for the instability in the nonlinear
regime, i.e., when the amplitudes are larger. We only need to multiply
$\Gamma_{\rm KH}$ by the corresponding ratio of amplitudes. An example is
represented in Fig.~\ref{gammakhi}, for a value of  $t/\tau_{\rm A}\approx 175$,
i.e., when the instability develops for the case  $\xi_0/R=0.2$, the value of
$\Gamma_{\rm KH}$ is around 1.5. Now we can use this value to infer the onset
time for other amplitudes. For the case $\xi_0/R=0.4$ the curve $\Gamma_{\rm
KH}$ intersects 1.5 (not shown here) at $t/\tau_{\rm A}\approx 90$ which is in
agreement with the results of the nonlinear simulations (see
Fig.~\ref{Fouriersum}). Therefore, Eq.~(\ref{eqgammakh}) can be used as a crude
estimate for the onset times of the KHI instability.

\citet{terradasetal2016} have used this assumption to estimate the onset times of
the instability in a curved flux rope configuration with an embedded prominence
and have found that slightly lower values for $\Gamma_{\rm KH}$ (around 0.6) are
required to have a good match with the simulations.

\subsubsection{Effect of numerical resolution}\label{sectresol}

\begin{figure}[!hh] \center{\includegraphics[width=7cm]{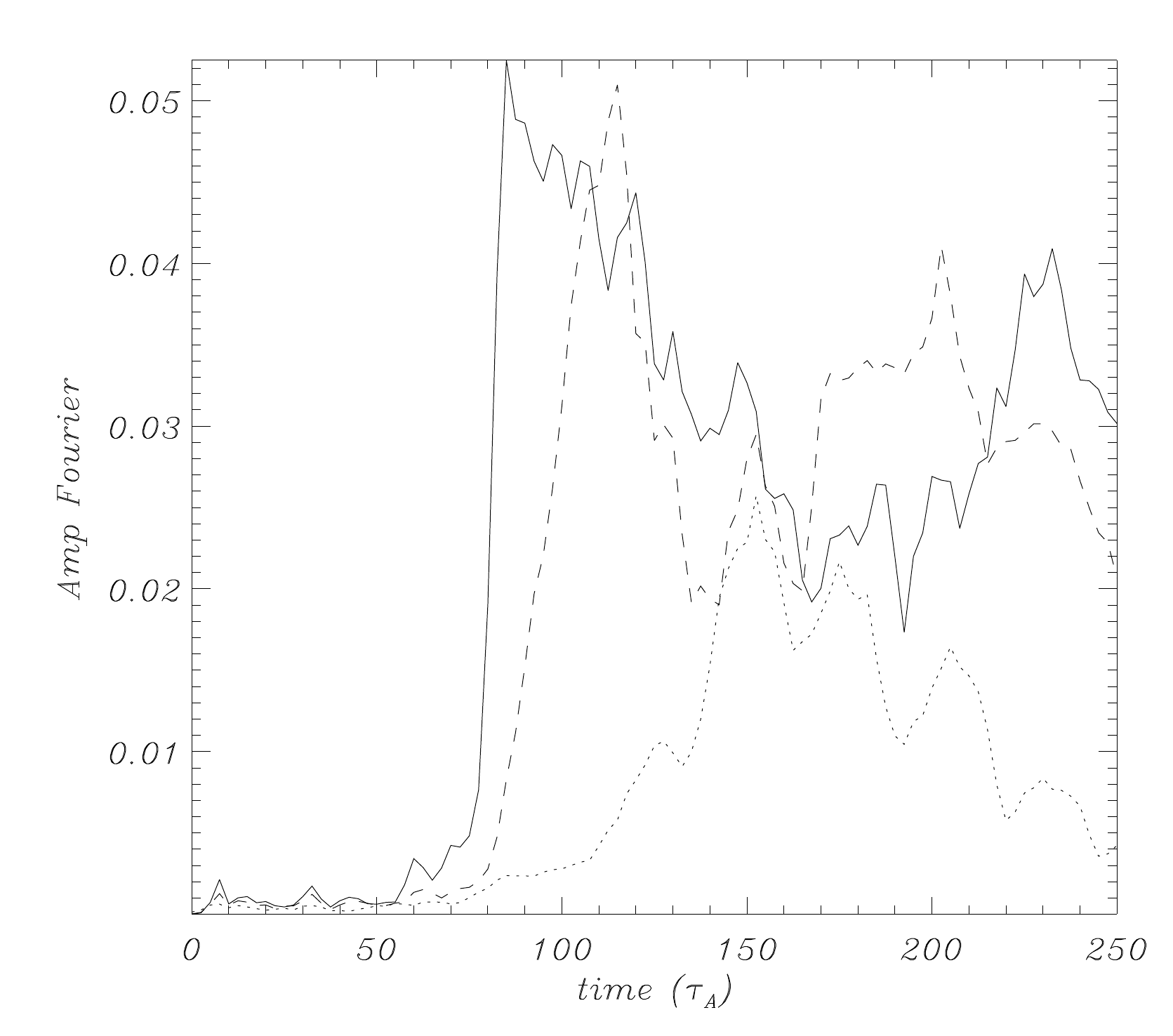}} 
\caption{\small Sum of Fourier coefficients,
$\sum_{p=3}^{N/2}\left|G(p)\right|$, as a function of time for the radial
velocity $v_{r R}$. The continuous, dashed, and dotted lines correspond to the
situation of high, medium an low resolution, respectively. In this case
$l/R=0.3$.}\label{sumfourtwistlow} \end{figure}

Contrary to the situation in the linear regime, in the nonlinear case the effect
of the mesh resolution is more determinant. The growth rates of the unstable KH
modes have a strong dependence on the azimuthal wavelengths of the modes.
Therefore it is expected that the shorter the azimuthal wavelength resolved by
the grid, the faster the onset of the instability. This is true when the results
for low and medium resolution are compared (see Fig.~\ref{sumfourtwistlow}).
Nevertheless, for medium and high resolution simulations, the times for the onset
of the instability are very similar, indicating convergence of the results. {\bf
We have also performed a simulation with [400, 400, 320] points, i.e., with
better resolution in the $z-$direction and have arrived to the same conclusion}.
Hence, we are sure that the numerical simulations are not significantly affected
by the number of mesh points as long as we use medium or high resolutions. The
same conclusion applies to the attenuation profiles.

\section{Results for the twisted tube}\label{twist}

The effect of magnetic twist on standing transverse oscillations is investigated
in this section. It is important to mention that the initial velocity
perturbation, which depends on the amount of twist, is crucial to avoid the
excitation of longitudinal harmonics. For example, an initial perturbation with
$n=1$ ($k_z=n\, \pi/L$) in $v_x$ and zero in $v_y$ significantly excites
the longitudinal mode $n=3$ in the case of a  twisted tube. The particular
perturbation given in Section \ref{eqsinit} allows us to concentrate on the
fundamental mode ($n=1$) and proceed as in the previous section.  

\subsection{Linear results}\label{lintwist}

In the linear regime the simulations show that the frequency of the standing
kink oscillation is essentially unaltered, meaning that $\omega=\omega_{\rm k}$.
This is in agreement with the analytical results of 
\citet{ruderterr2015} in the situation of weak twist and for
exactly the same magnetic twist model \citep[see also][for other twist
profiles]{ruderman07}. The effect of twist on propagating waves is much more
significant, for example regarding the change in the kink frequency, e.g., 
\citet{terradasgoossens2012,ruderman2015}.

\begin{figure}[!hh] \center{\includegraphics[width=7cm]{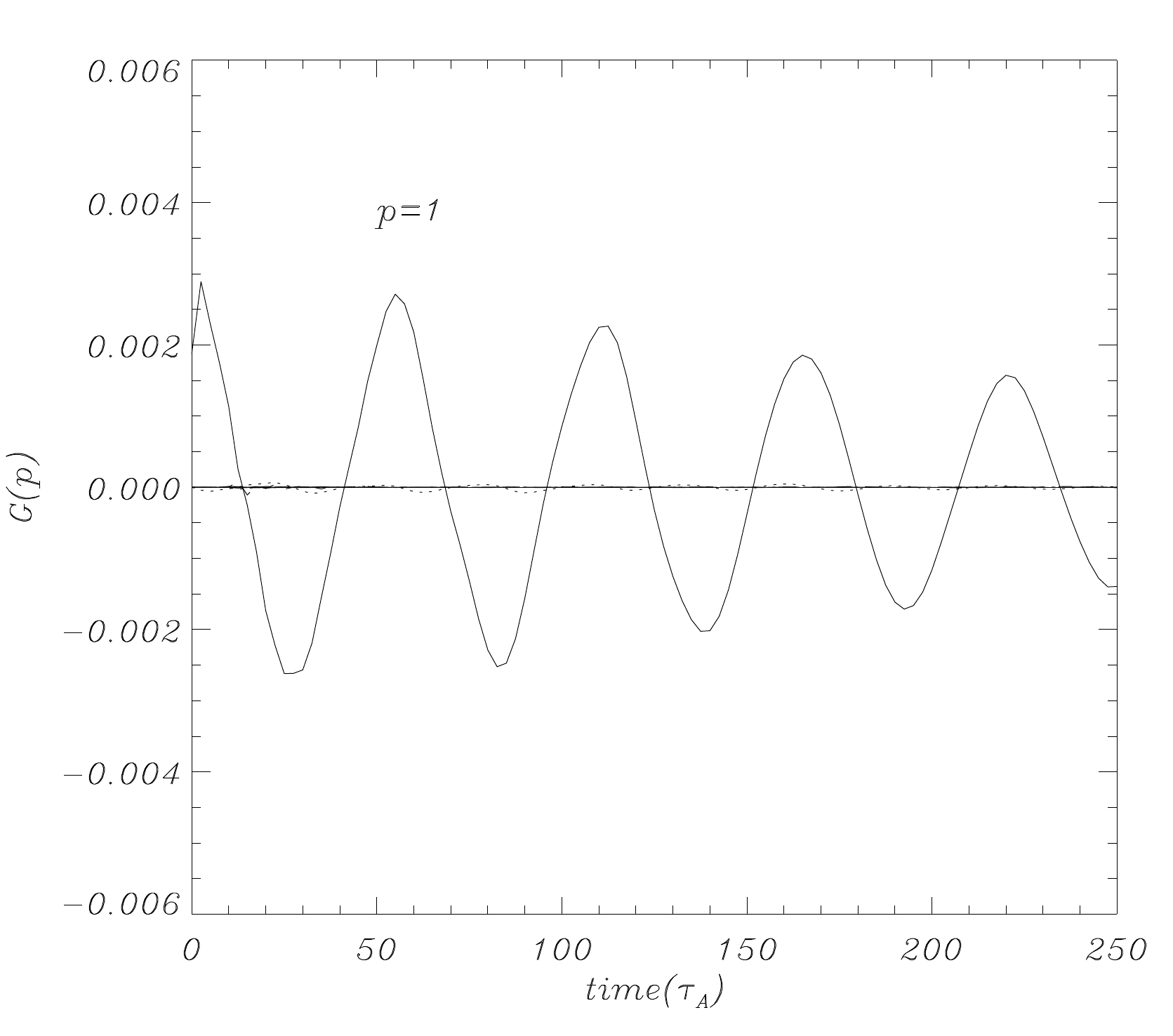}}
\center{\includegraphics[width=7cm]{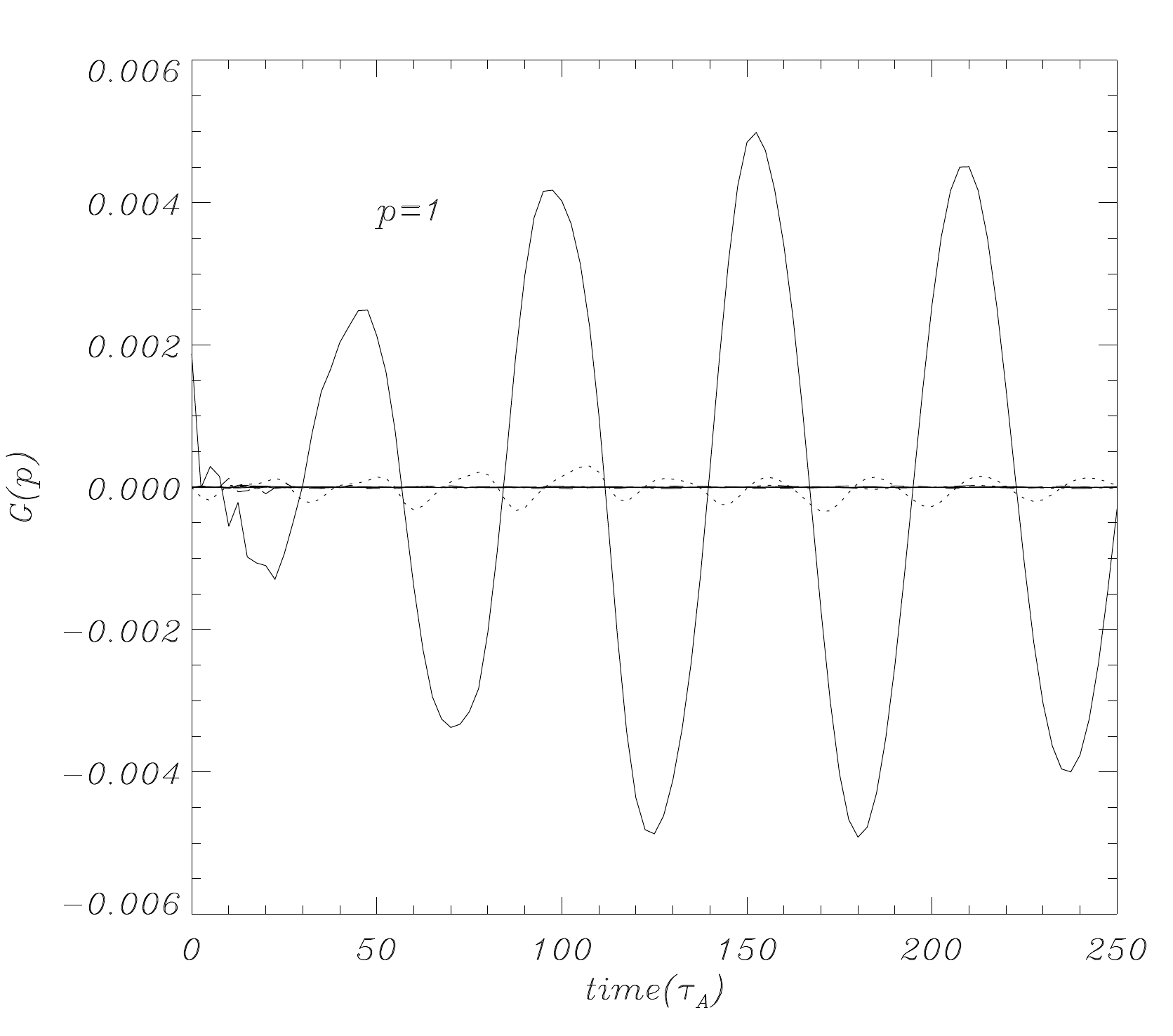}} \caption{\small Fourier
coefficients as a function of time for the radial velocity $v_{r R}$ (top panel)
and azimuthal velocity $v_{\phi R}$ (bottom panel) in the linear regime. The
contribution of the $p=1$ is dominant. In this case $l/R=0.3$ and
$B_\phi/B_z=0.2$.}\label{Fouriercoeftw0} \end{figure}

For standing waves the polarization of the motions along the loop axis changes
according to the profile derived by \citet{ruderterr2015} and given by
Eqs.~(\ref{twistpol1})-(\ref{twistpol}) \citep[see
also][]{terradasgoossens2012}. Regarding the attenuation associated to the
inhomogeneous layer plus magnetic twist, so far there are no studies about the
damping times for standing waves in such a model, although some work has been
done for propagating waves \citep[see][]{karamibaha10}. A detailed eigenmode
analysis is required but it is out of the scope of this work since the main aim
here is the KHI. It suffices to mention that from the time-dependent simulations
we do not find a significant change in the efficiency of the resonant damping
mechanism for the values of twist considered in this work.

In  Fig.~\ref{Fouriercoeftw0} (top panel) an example of the time evolution of the
Fourier coefficients is displayed. Again $p=1$ is dominant and the excitation of
$p=2$ is extremely weak since we are in the linear regime. In this situation, an
increase on the grid resolution has the same effect as in the non-twisted loop. The
damping times are not affected, but short spatial scales, taking place around the
resonant positions, are resolved better. In  Fig.~\ref{Fouriercoeftw0} (bottom
panel) we find again the increase of amplitude of the azimuthal component
associated to the resonant damping. In this plot $p=2$ can be also identified.

It is worth to mention that, in the case of a twisted magnetic tube, it is more
convenient to use instead of the azimuthal velocity component, $v_{\phi}$, the
perpendicular component to the magnetic field lying on magnetic surfaces (still
cylindrical) and defined as $v_{\perp}=\left(B_z v_{\phi}-B_{\phi}
v_z\right)/B$. Twist introduces a vertical component of the velocity, $v_z$,
which in our case has other contributions due to gas pressure and ponderomotive
forces. It is also appropriate to use the parallel component of the velocity to
the magnetic field lines, $v_{\parallel}=\left(B_{\phi}v_{\phi}+B_z 
v_z\right)/B$. This component of the velocity is strictly zero only when the
plasma-$\beta$ is zero and when the linearized MHD equations are considered.

\begin{figure}[!hh] \center{\includegraphics[width=7cm]{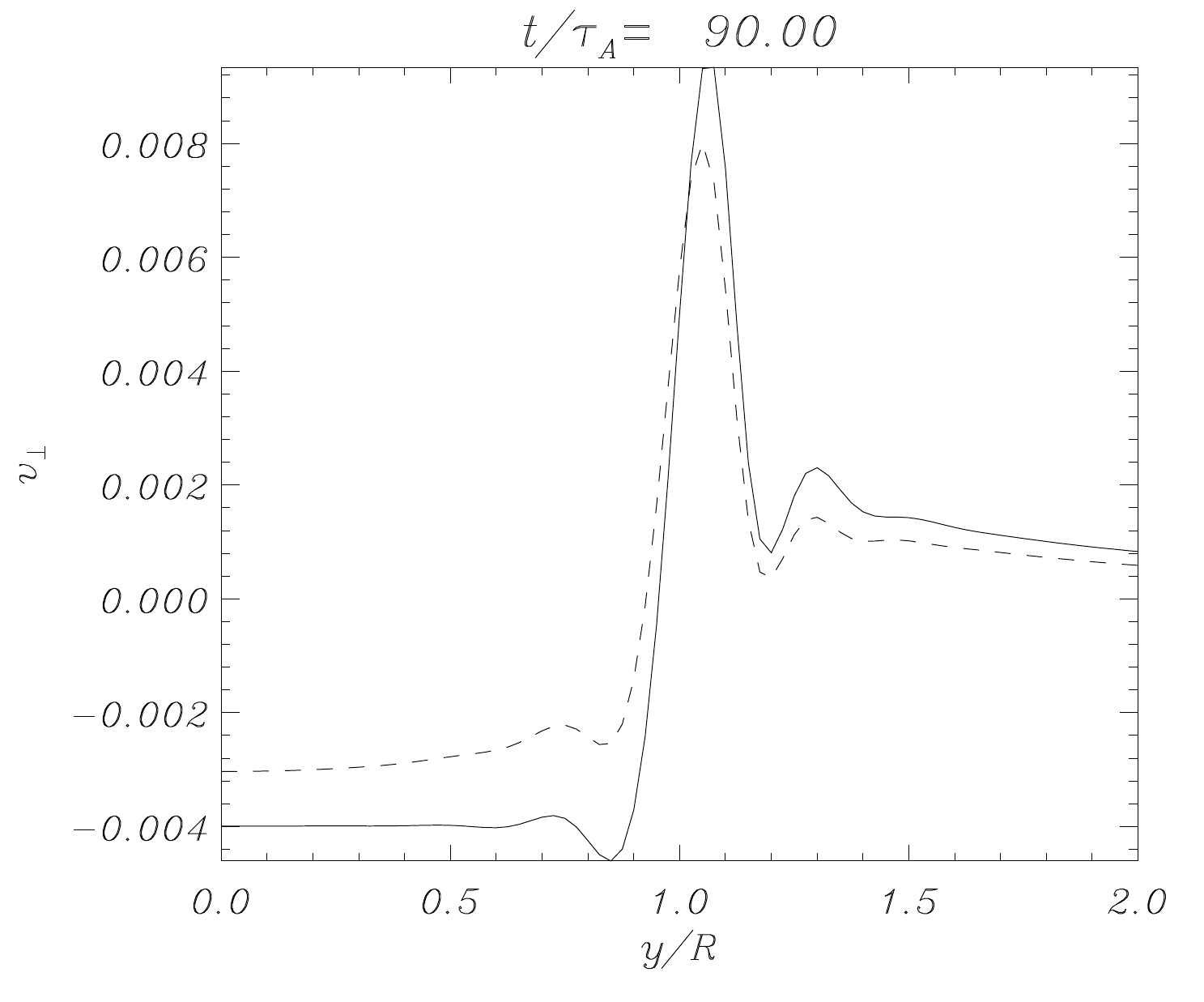}}
\center{\includegraphics[width=7cm]{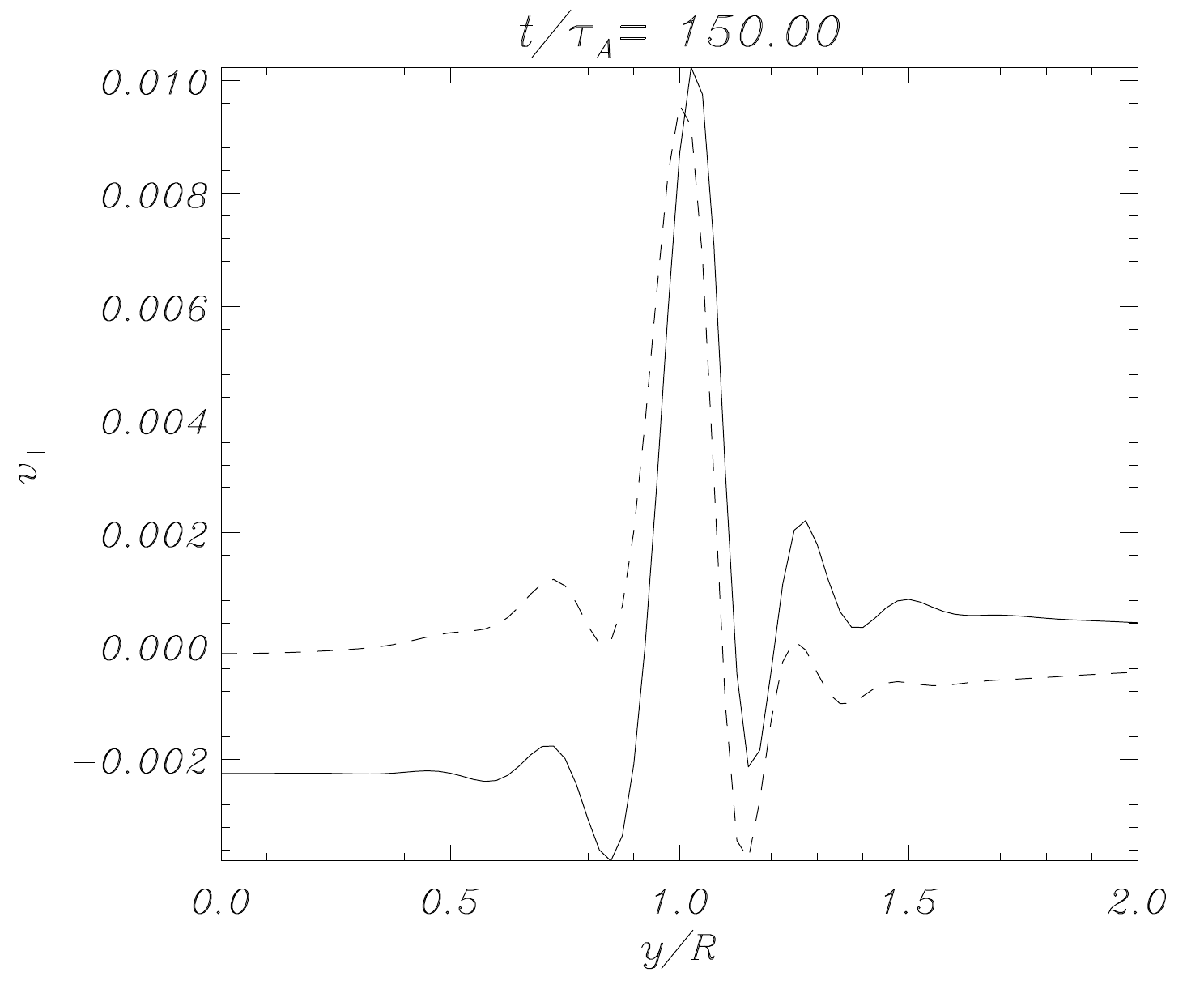}} \caption{\small Perpendicular
velocity component to the magnetic field, $v_{\perp}$, as a function of position
(in a range between the center of the tube and $2R$) at two different times. The
continuous line corresponds to the case without twist while the dashed line
corresponds to the situation $B_\phi/B_z=0.4$. In these simulations $l/R=0.3$
and the linear regime is considered.}\label{vperplayer} \end{figure}

\begin{figure}[!hh] \center{\includegraphics[width=7cm]{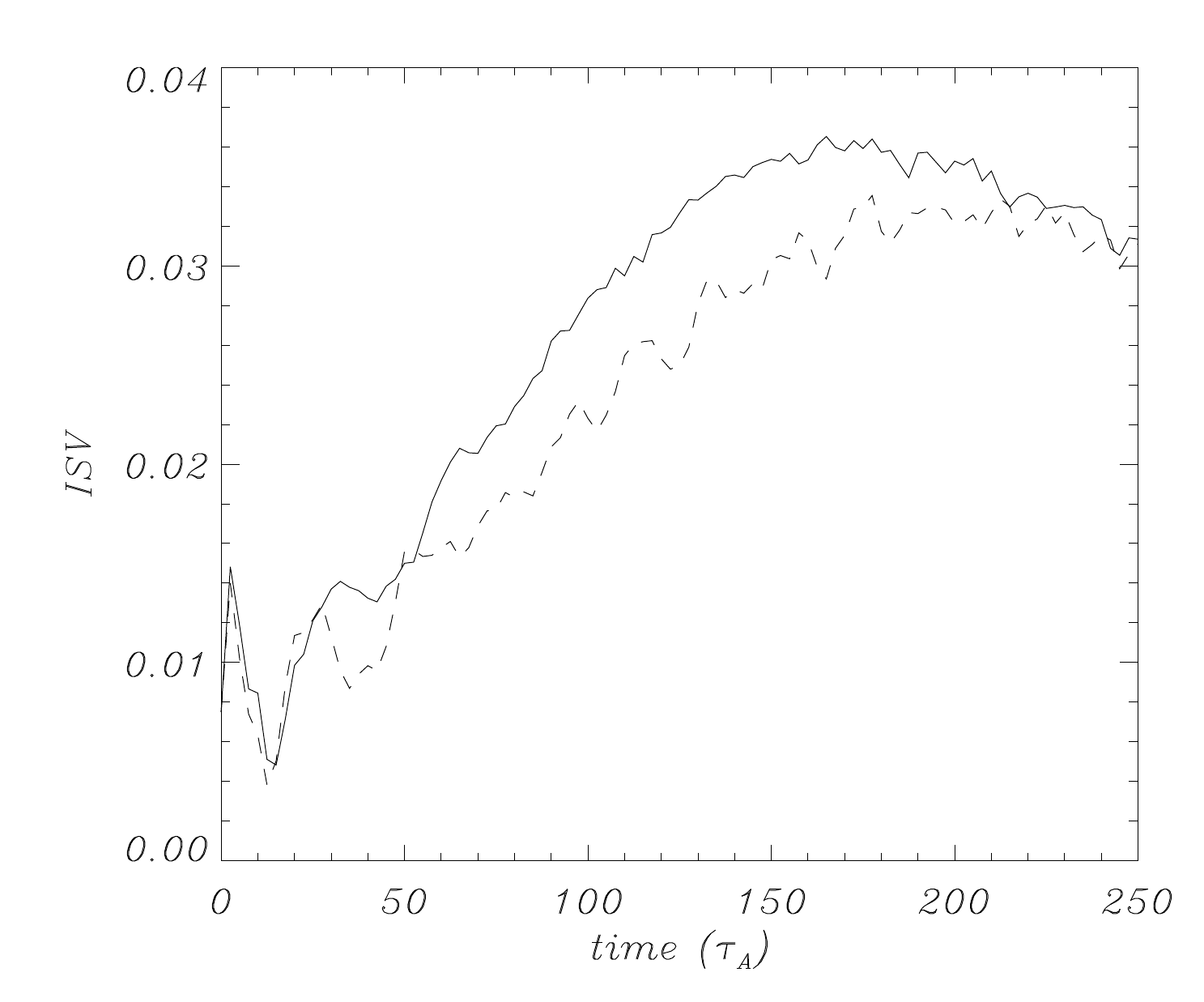}}
\caption{\small Integrated shear vorticity (ISV) as a function of time. The
continuous line corresponds to the case without twist while the dashed line
represents the situation for $B_\phi/B_z=0.4$.}\label{ivs} \end{figure}

An example of the shear in $v_\perp$ and in the linear regime is plotted in
Fig.~\ref{vperplayer} across the inhomogeneous layer. For comparison purposes
the result with magnetic twist (dashed line) is plotted together with the
simulation without twist (continuous line). The peak around $y/R=1$ is due to
the energy transference between the global motion of the tube and the Alfv\'enic
localized oscillations. The spatial scales in the layer decrease with time, and
consequently the shear increases (compare top and bottom panel) due to the phase
mixing process. Interestingly, the results shown in Fig.~\ref{vperplayer} reveal
that the motions for the tube with magnetic twist seem to have less shear than
the motions for the untwisted tube. To properly quantify this effect, we have
computed the shear vorticity, defined as the derivative of $v_{\perp}$ along the
$y-$direction. The idea is to integrate this magnitude in space ($0<y/R<2$) but
taking its absolute value to avoid effects due to changes in sign. This provides
a proxy for the total shear associated to the oscillations of the tube. The
integrated shear vorticity (ISV) is plotted in  Fig.~\ref{ivs} as a function of
time. The interpretation is the following, at $t=0$ the ISV is essentially the
same for the untwisted and twisted tubes and starts to fluctuate due to the kink
motion. Before completing a single period the ISV increases due to the
generation of small scales and the rise in $v_{\perp}$ because energy is pumped
into the layer. Then it reaches a saturation point where numerical dissipation
prevents a further decrease of the spatial scales. But the most relevant
information that we can extract from Fig.~\ref{ivs} is that, in the case of a
twisted tube, the ISV is, after a short transient, always below the value for the
untwisted tube and its growth-rate is smaller. This means that one of the effects
of twist is to reduce the total shear velocity at the tube boundary with respect
to the oscillations in the untwisted tube. As we show in the following
sections this has important consequences regarding the development of the KHI. 

\begin{figure}[!hh] \center{\includegraphics[width=7cm]{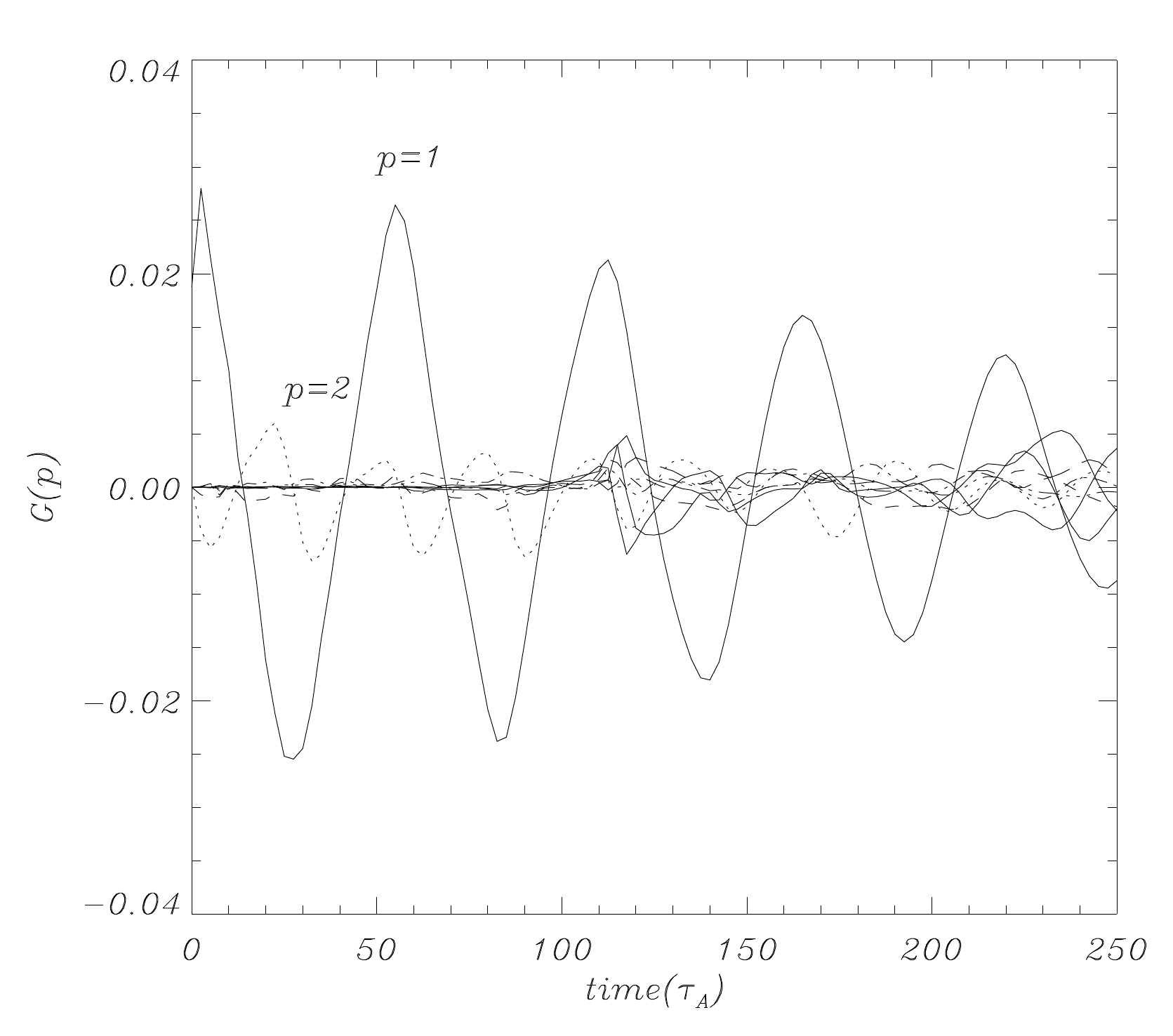}}
\center{\includegraphics[width=7cm]{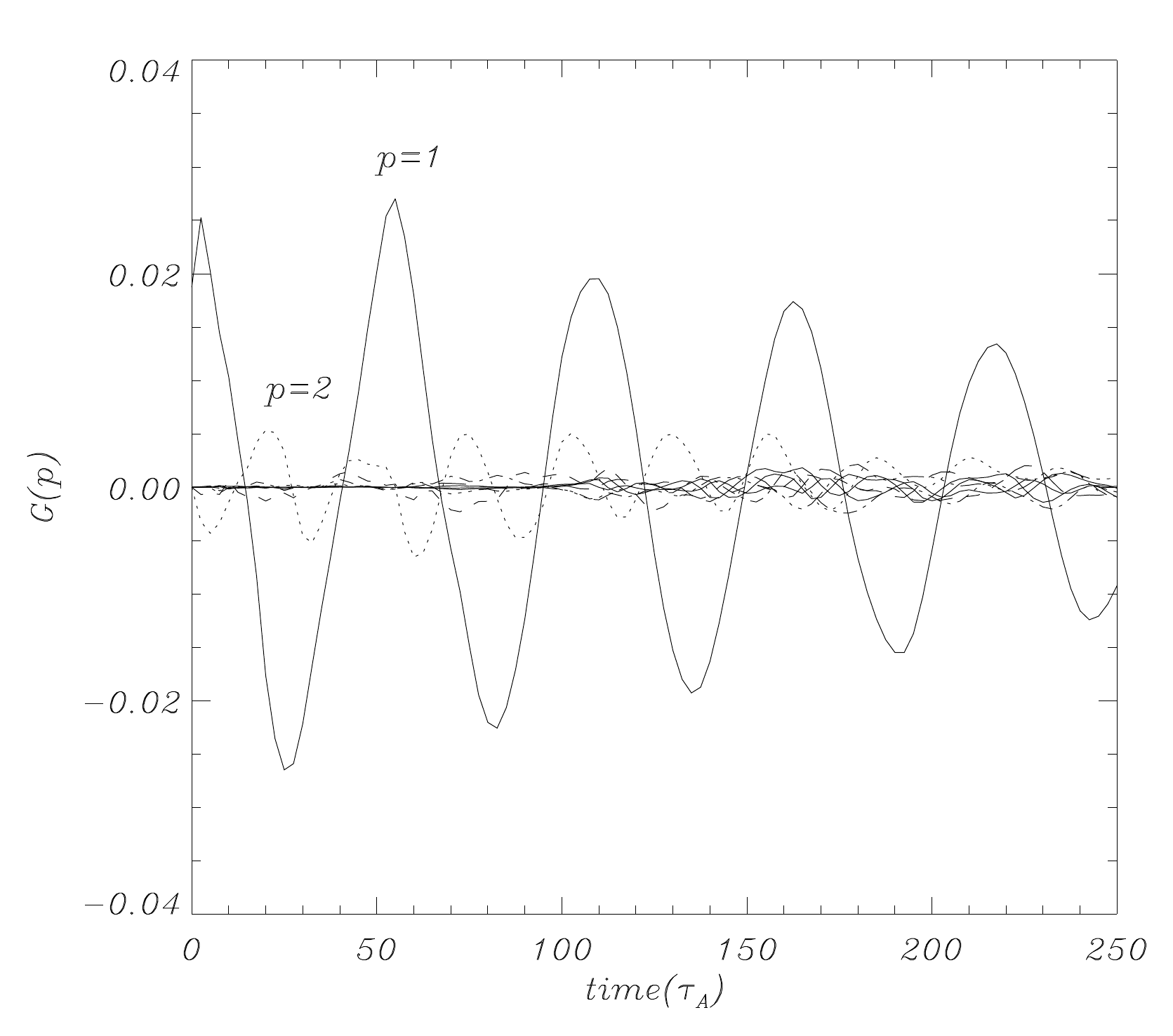}} \caption{\small Fourier
coefficients as a function of time for the radial velocity $v_{r R}$. The
contribution of the $p=1$ is dominant. Higher values $p$s are due to nonlinear
effects. The $p=2$ is oscillating at half the period of $p=1$. In this case
$l/R=0.3$. In the top panel $B_\phi/B_z=0.2$ while in the bottom panel
$B_\phi/B_z=0.4$. The excitation is nonlinear
($\xi_0/R=0.4$).}\label{Fouriercoeftw} \end{figure}

\begin{figure}[!hh] \center{\includegraphics[width=7cm]{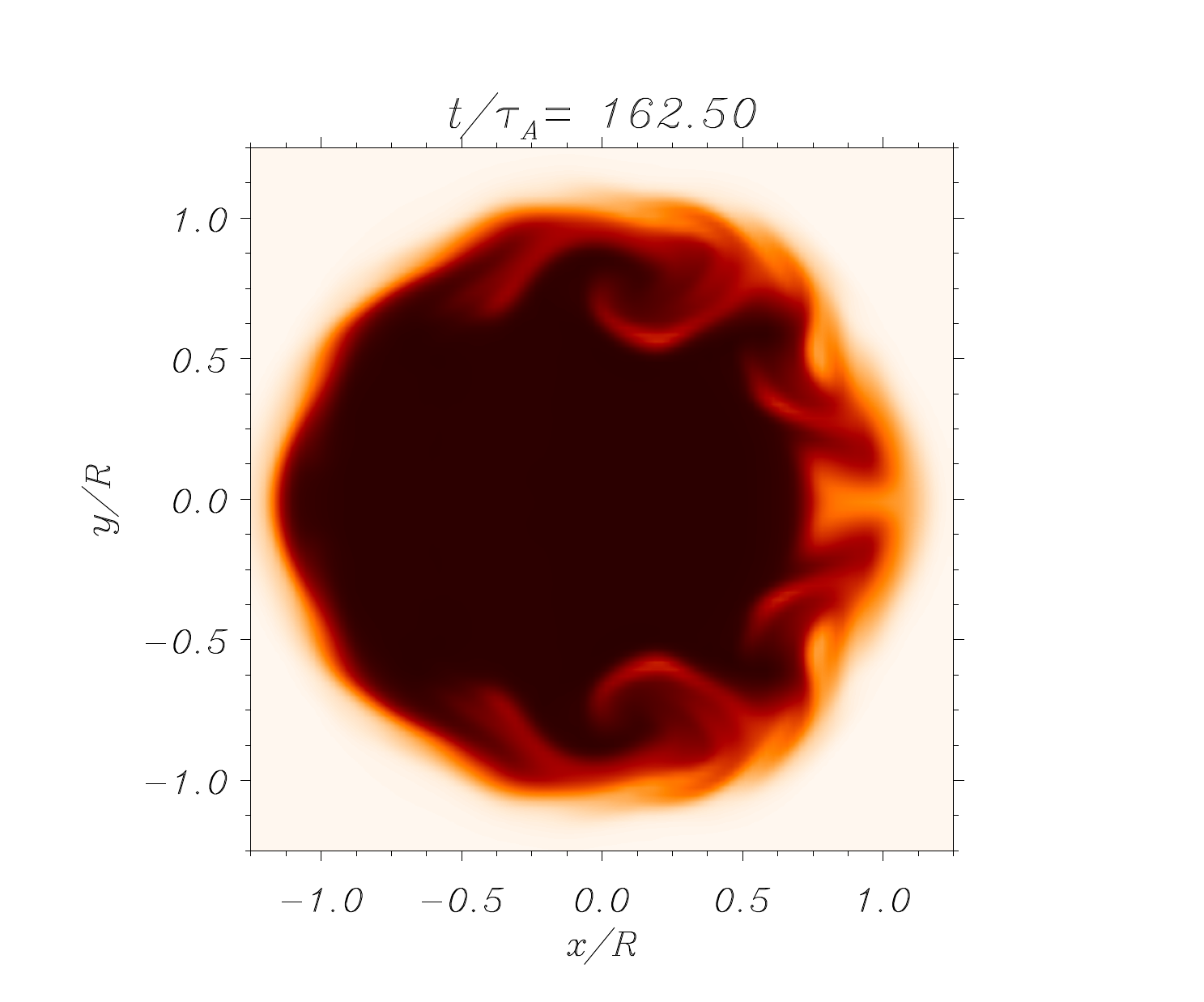}}
\center{\includegraphics[width=7cm]{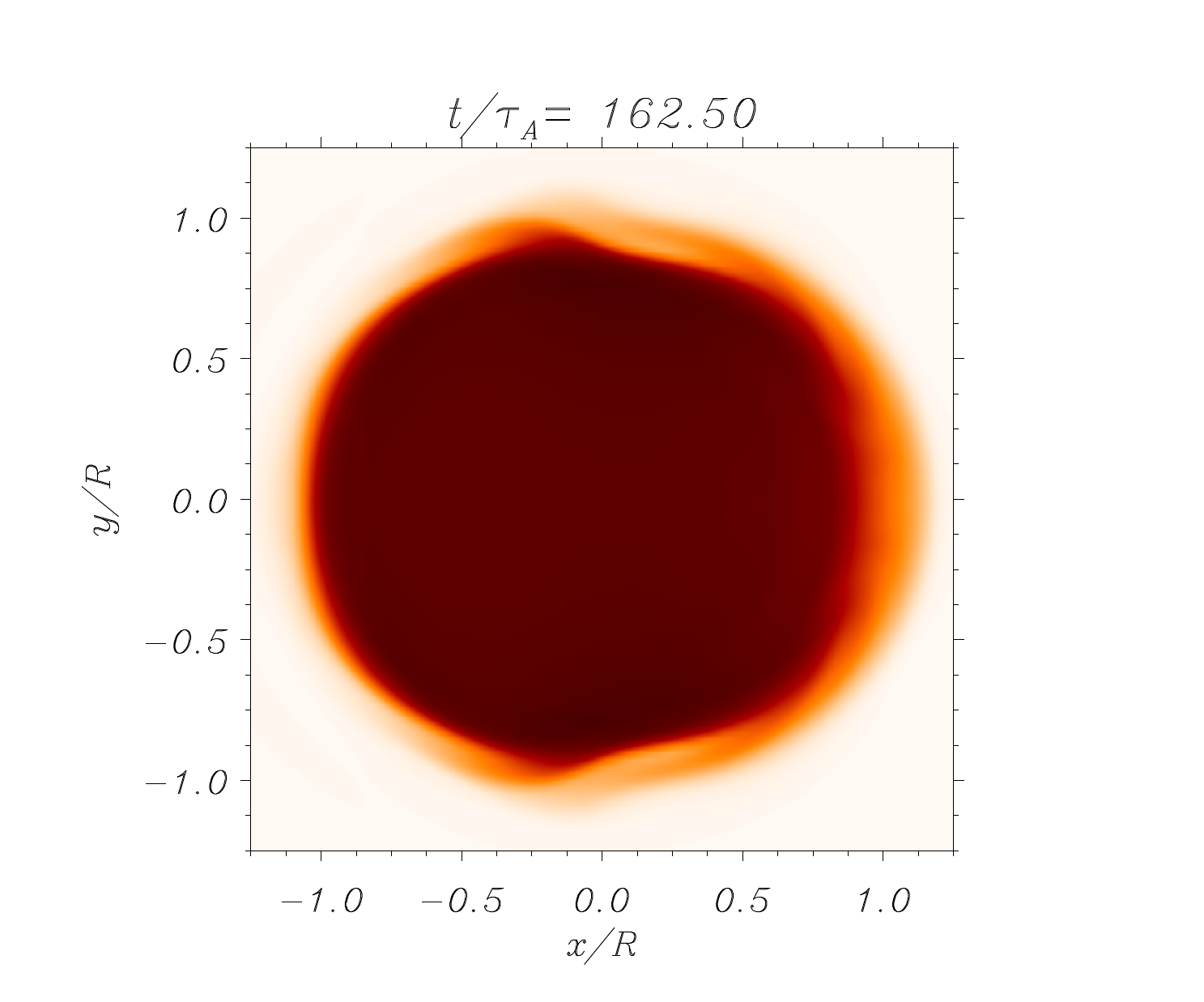}} \caption{\small Snapshot of the
two-dimensional density distribution at half the tube length ($z=L/2$) at a given
time. In the top panel $B_\phi/B_z=0.2$ while in the bottom panel $B_\phi/B_z=0.4$.
The development of the KHI is delayed when magnetic twist is increased. For this
simulations $l/R=0.3$.}\label{denstwist} \end{figure}

\subsection{Nonlinear results}\label{nonlintwist}

In the nonlinear regime ($\xi_0/R=0.4$) and under the presence of twist, we again
find the development of the KHI. However, now the azimuthal component of the
magnetic field changes the properties of the oscillation. Since the polarization of
the motion varies along the tube, the shear flow is less organized in the twisted
case. The shear flow profile due to $v_{\perp}$ at the tube boundary is still
responsible for the instability and by definition this velocity component is 
perpendicular to the equilibrium magnetic field. Therefore, the equilibrium
magnetic field should not have the expected stabilizing effect of the situation
when there is a magnetic component along the shear flow \citep[e.g.,][]{chandra61}.
Nevertheless, twist changes the properties of the transverse oscillation, as we
have already seen in the linear regime, and has an effect on the instability onset.

\begin{figure}[!hh] 
\center{\includegraphics[width=7cm]{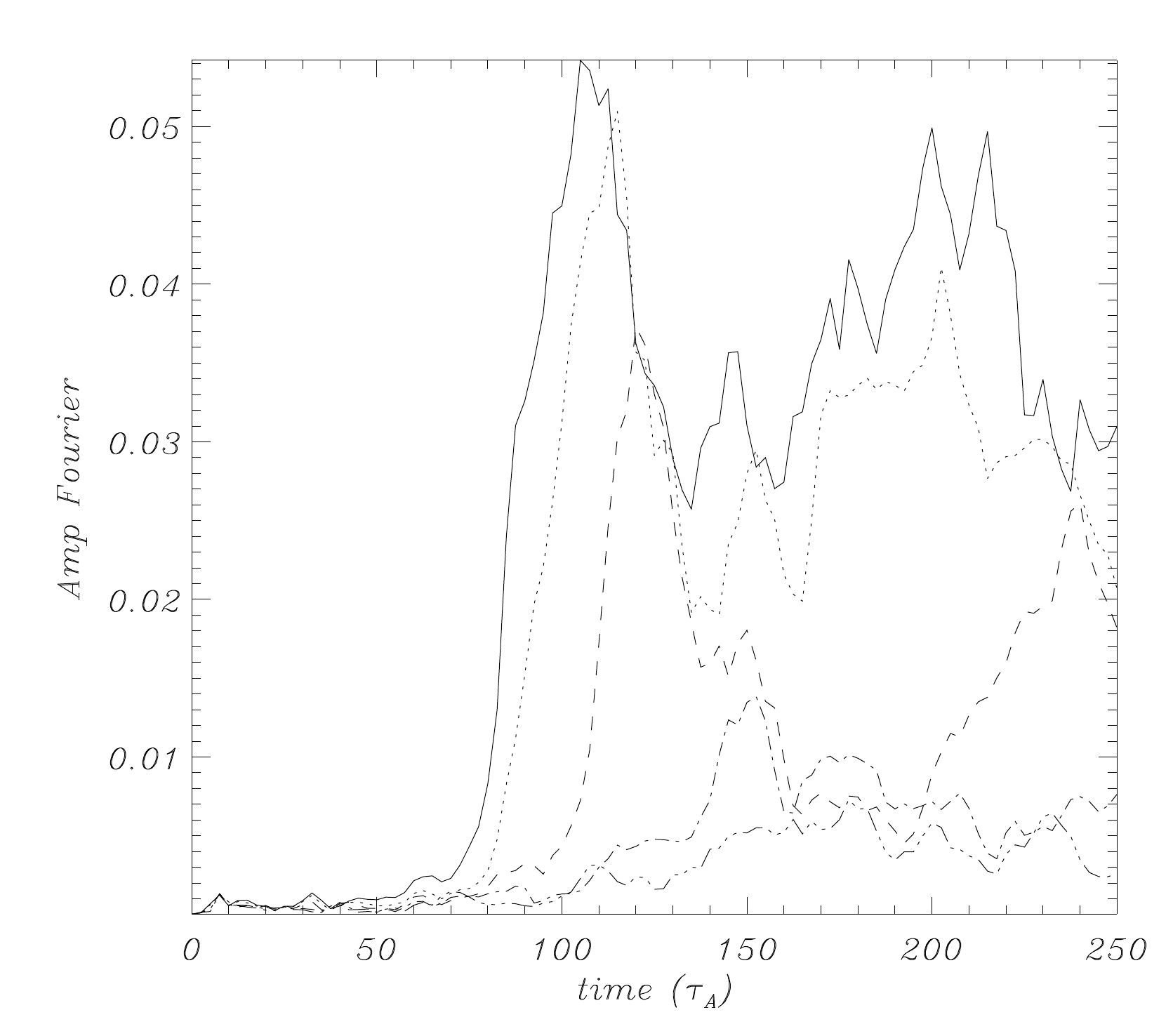}} \caption{\small
Sum of Fourier coefficients, $\sum_{p=3}^{N/2}\left|G(p)\right|$, as a
function of time for the radial velocity $v_{r R}$. The continuous, dotted,
dashed, dot$-$dashed and triple dot$-$dashed lines correspond to the cases
with $B_\phi/B_z=0,0.1, 0.2, 0.3$ and $0.4$, respectively. In this case
$l/R=0.3$.}\label{sumfourtwist} \end{figure}

To facilitate the comparison of the results we proceed as in
Section~\ref{nonlinnotwist} and concentrate on the plane at $z=L/2$. We have
calculated the Fourier coefficients for two different values of twist
($B_\phi/B_z=0.2$ and 0.4), which are plotted in Fig.~\ref{Fouriercoeftw}. We
can compare these plots with the results for the same situation without twist,
represented in bottom panel of Fig.~\ref{Fouriercoef}. The coupling with $p=2$
is still present and the amplitude of this mode does not seem to depend strongly
on the amount of twist. Nevertheless, the main differences between the plots are
in the timing for the onset of the excitation of higher order values of $p$.
Further evidence of this effect is found in the density distribution for the two
twisted tube models at the same instant, plotted in Fig.~\ref{denstwist}. The
deformations of the boundary are less pronounced in the case of $B_\phi/B_z=0.4$
in comparison with the situation $B_\phi/B_z=0.2$. These plots can be also
compared with the untwisted tube, shown in bottom panel of
Fig.~\ref{densnotwist}, displaying a much more complex structure in the
azimuthal direction. It is clear that magnetic twist delays the appearance of
the KHI instability. However, we have to bear in mind that the helical vortices
introduce structuring along the $z-$direction and that these vortices may end
up  colliding into each other due to the different growth rates depending on the
direction of swing. This can significantly change the structuring in the
azimuthal direction at more evolved stages of the system.

\begin{figure}[!hh]  \center{\includegraphics[width=7cm]{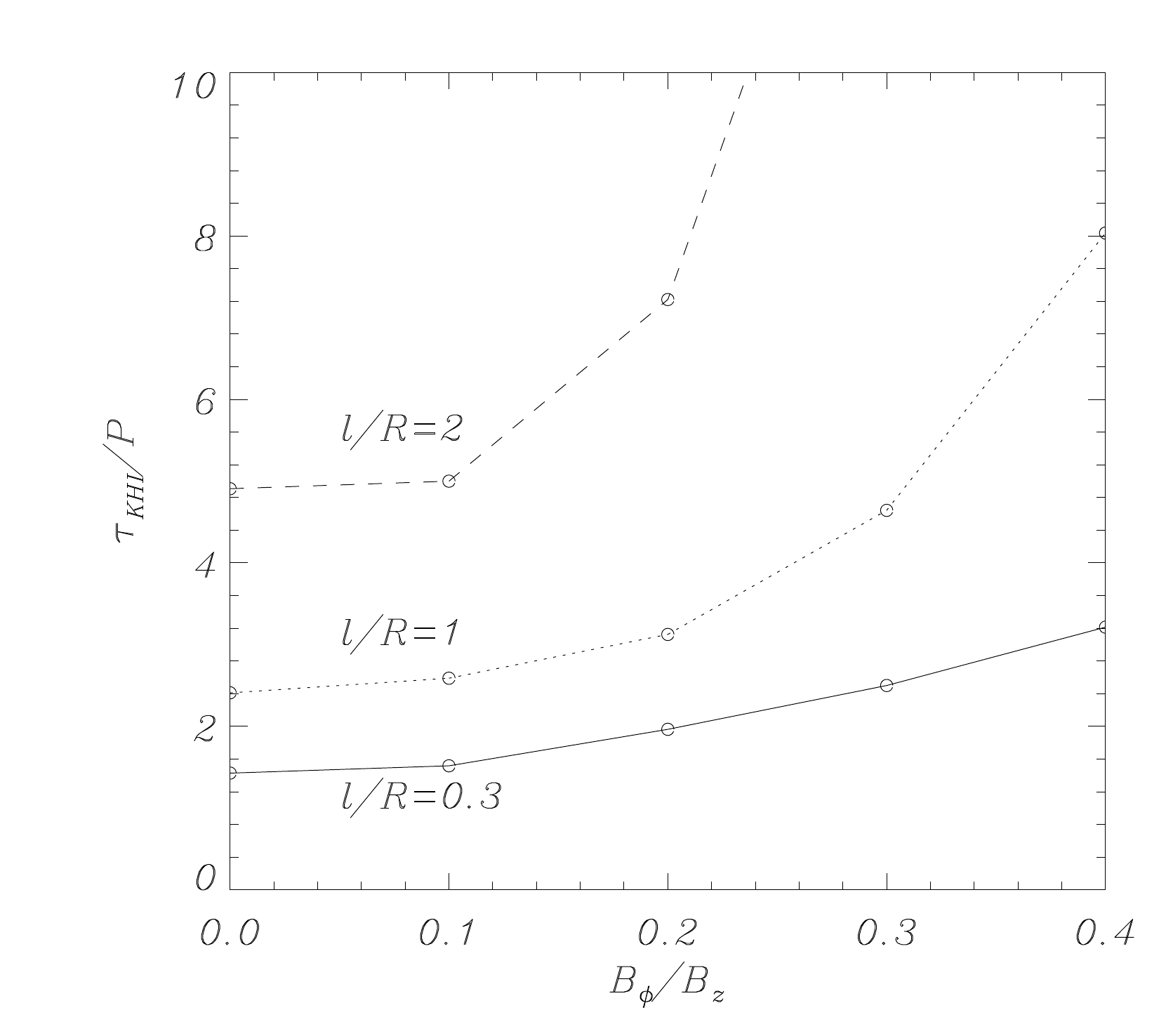}}
\caption{\small Onset time of the KHI normalized to the kink period as a
function of twist calculated from Fig.~\ref{sumfourtwist} for an inhomogeneous
layer of $l/R=0.3$. The results for thicker layers, $l/R=1$ and $l/R=2$, are
also shown.}\label{tauinst} \end{figure}

In Fig.~\ref{sumfourtwist} the contribution of high order $p$s is plotted as a
function of time for different values of twist. The stronger the azimuthal
component of the field, the later the development of the KHI. From
Fig.~\ref{sumfourtwist}, we have estimated the onset time for the instability,
$\tau_{\rm KHI}$, according to a given amplitude threshold, in this case 0.01. The
results, shown in Fig.~\ref{tauinst}, indicate that the slope of the curve
increases when twist is augmented. A twisted tube with $B_\phi/B_z=0.4$ and
$l/R=0.3$ has a typical onset time for the instability around three times larger
than the untwisted tube.

\begin{figure}[!hh] \center{\includegraphics[width=7cm]{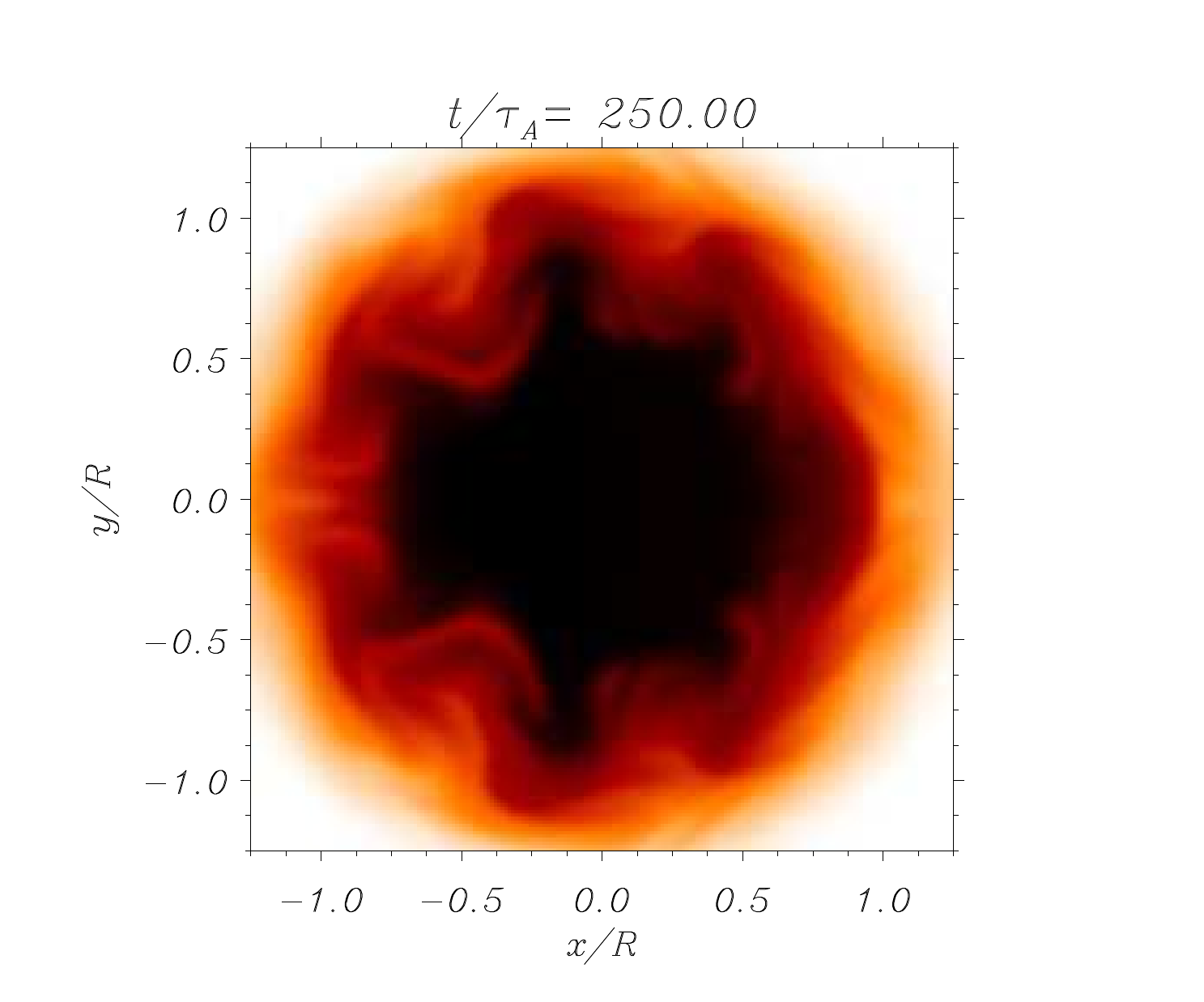}}
\center{\includegraphics[width=7cm]{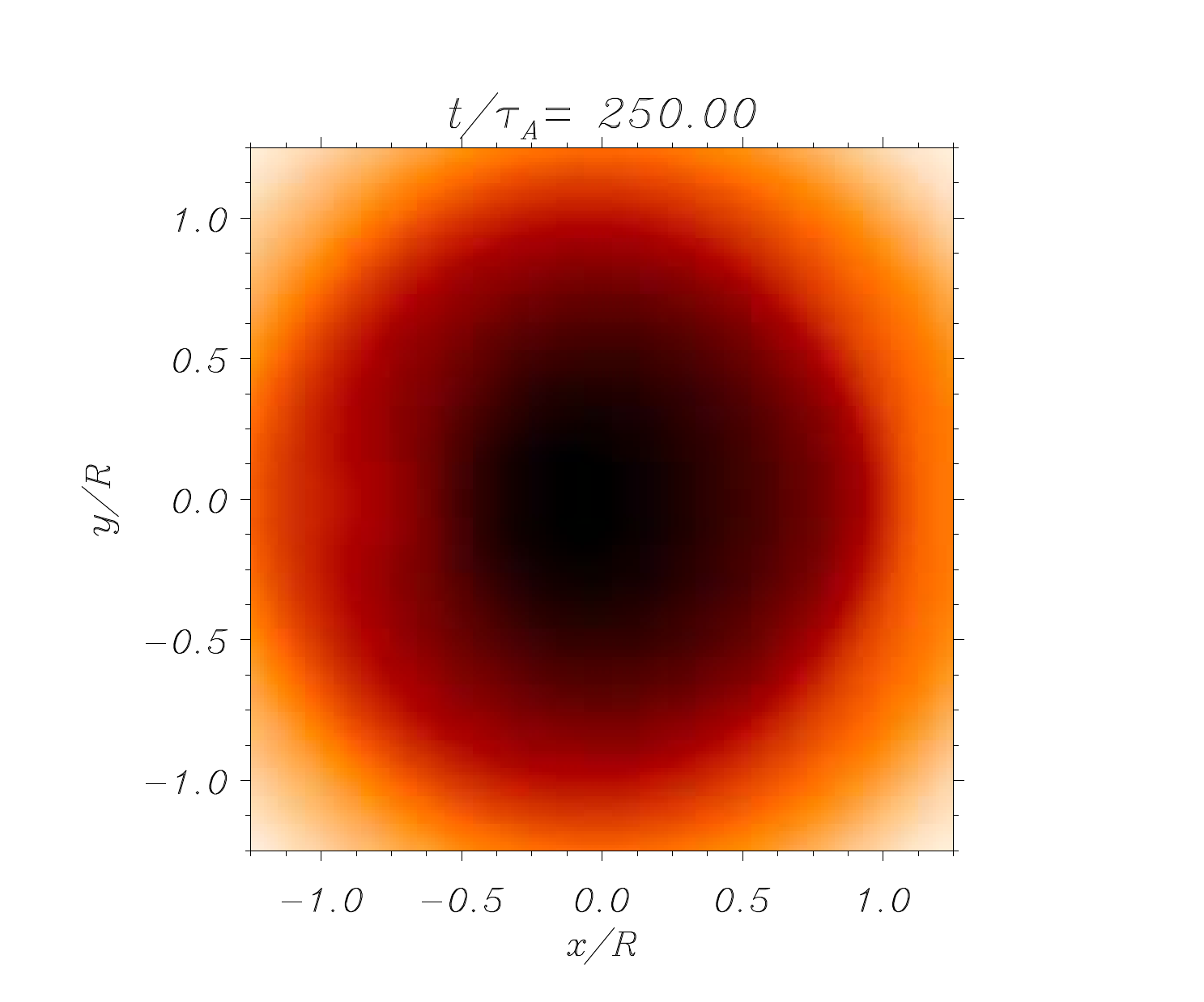}} \caption{\small Snapshot of the
two-dimensional density distribution at half the tube length ($z=L/2$) at a given
time. In the top panel $l/R=1$ while in the bottom panel $l/R=2$ (fully
inhomogeneous loop). For these simulations $B_\phi/B_z=0.2$.}\label{denstwistl}
\end{figure}

So far, we have concentrated on a given width of the inhomogeneous layer
($l/R=0.3$), but it is also interesting to repeat the simulations changing $l/R$
to understand how the onset times of the instability are modified. The results
are also plotted in Fig.~\ref{tauinst}. We find that, as the thickness of the
layer is increased, the instability appears at later times, see the density
distribution of Fig.~\ref{denstwistl} for a particular case. This behavior is
already found in the untwisted tube. The interpretation is related to the
generation of the spatial scales at the inhomogeneous layer due to the
phase-mixing process. Since the generation of the velocity shear scales as
$L_{\rm ph}\approx l /(\Omega \,t)$, i.e., approximately linear in $l$, it means
that for thick layers the time required to achieve a given spatial scale is
always larger than for thin layers. Hence, following this argument, the
development of the instability should be always faster when $l/R$ decreases.
However, the shear length-scales are not the only factor that determines the
growth of the instability, the amplitude of the shear is also relevant. This
amplitude depends on the energy transference to the inhomogeneous layer, and it
grows faster when $l/R$ increases. The two effects are superimposed but from our
results it seems that the generation of small length-scales dominates, since we
always find that the instability develops faster when $l/R$ decreases. This
behavior is enhanced by the effect of twist. From Fig.~\ref{tauinst} we also find
that the dependence of the onset of the instability with twist is stronger for
thick layers (compare the curves for $l/R=0.3$ and $l/R=2$). Therefore, the
combination of significantly wide layers and strong twist produce the most stable
configurations regarding the KHI. 

\begin{figure}[!hh]  \center{\includegraphics[width=7cm]{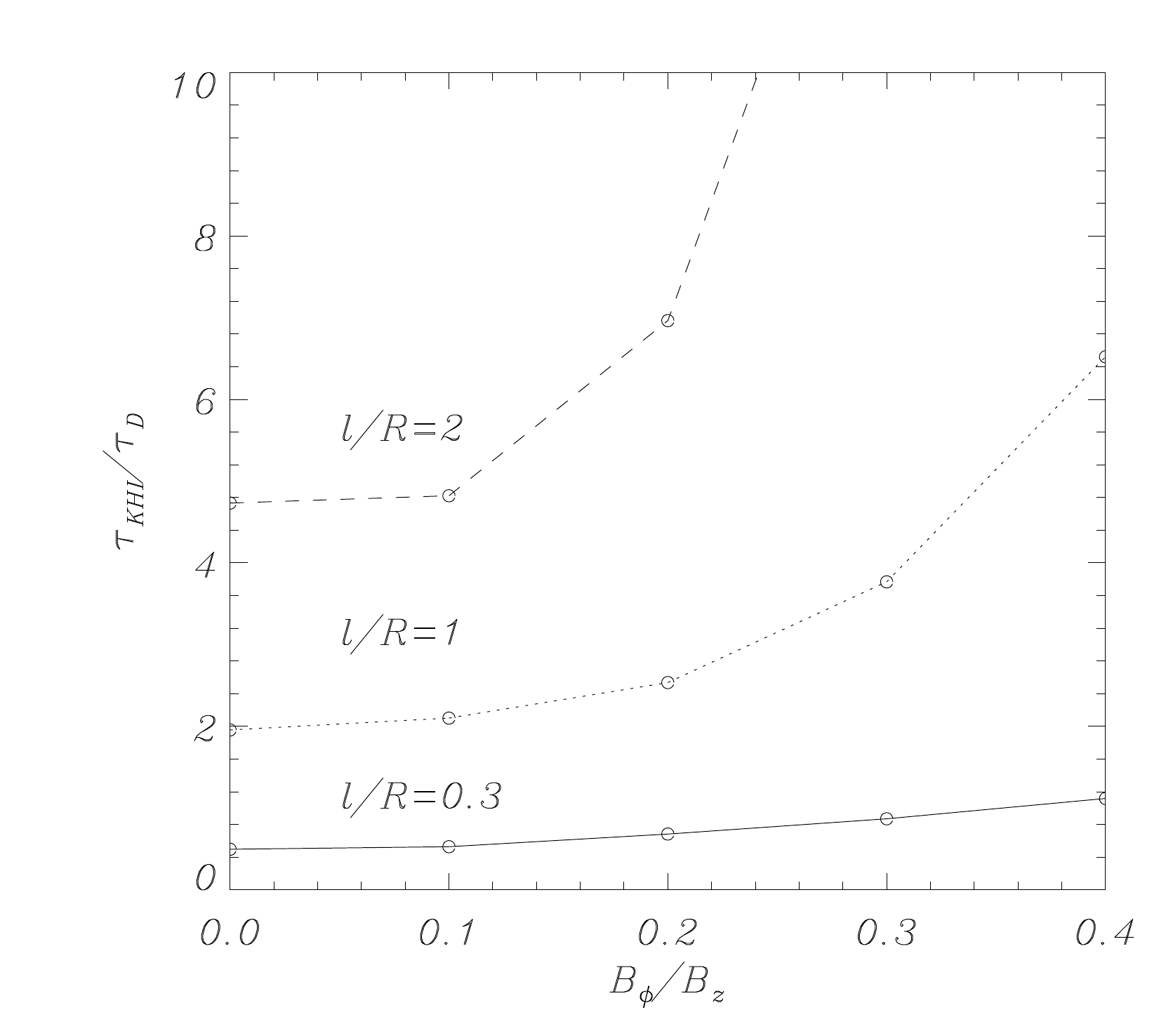}}
\caption{\small Ratio of KHI onset time to the damping time  as a function of
twist for three different widths of the inhomogeneous layer. The same parameters
as in Fig.~\ref{tauinst} are used.}\label{tauinsttaud} \end{figure}

However, we have to bear in mind that tubes with thick layers have much shorter
damping times than for the case with thin layers, and the global displacement of
the tube  is strongly attenuated after a few periods of oscillation. Under such
circumstances the instability will develop after the global oscillation has died
away. For this reason it is also interesting to compare the time-scales of the
attenuation, i.e., the damping time, $\tau_{\rm D}$, with the onset times for the
KHI. The results are displayed in Fig.~\ref{tauinsttaud}. From this figure we
conclude that only for thin layers the development of the instability can be
expected while the tube is clearly oscillating laterally (i.e., the situation
$\tau_{\rm KHI}/\tau_{\rm D}<1$), while for the case of thick layers and strong
twist the development of the instability takes place after the global transverse
oscillation has a negligible amplitude. Note that these results depend on the
amplitude of the initial excitation ($\xi_0/R=0.4$ in the present case) but the
chosen value can be viewed as representative of typical transverse loop
oscillations although larger values are usually reported \citep[see][for a
detailed study]{goddardnaka2016}.

Unfortunately, it has not been possible to infer the onset times for the
instability using the semianalytic approach used in Section~\ref{ankhi} for
the untwisted tube. The reason is that the presence of the magnetic field
changes the nature of the Rayleigh equation that has to be solved and the
methods used by \citet{heypri83} and \citet{brownpri84} are not easily extended
to this situation.

\begin{figure}[!hh]  \center{\includegraphics[width=7cm]{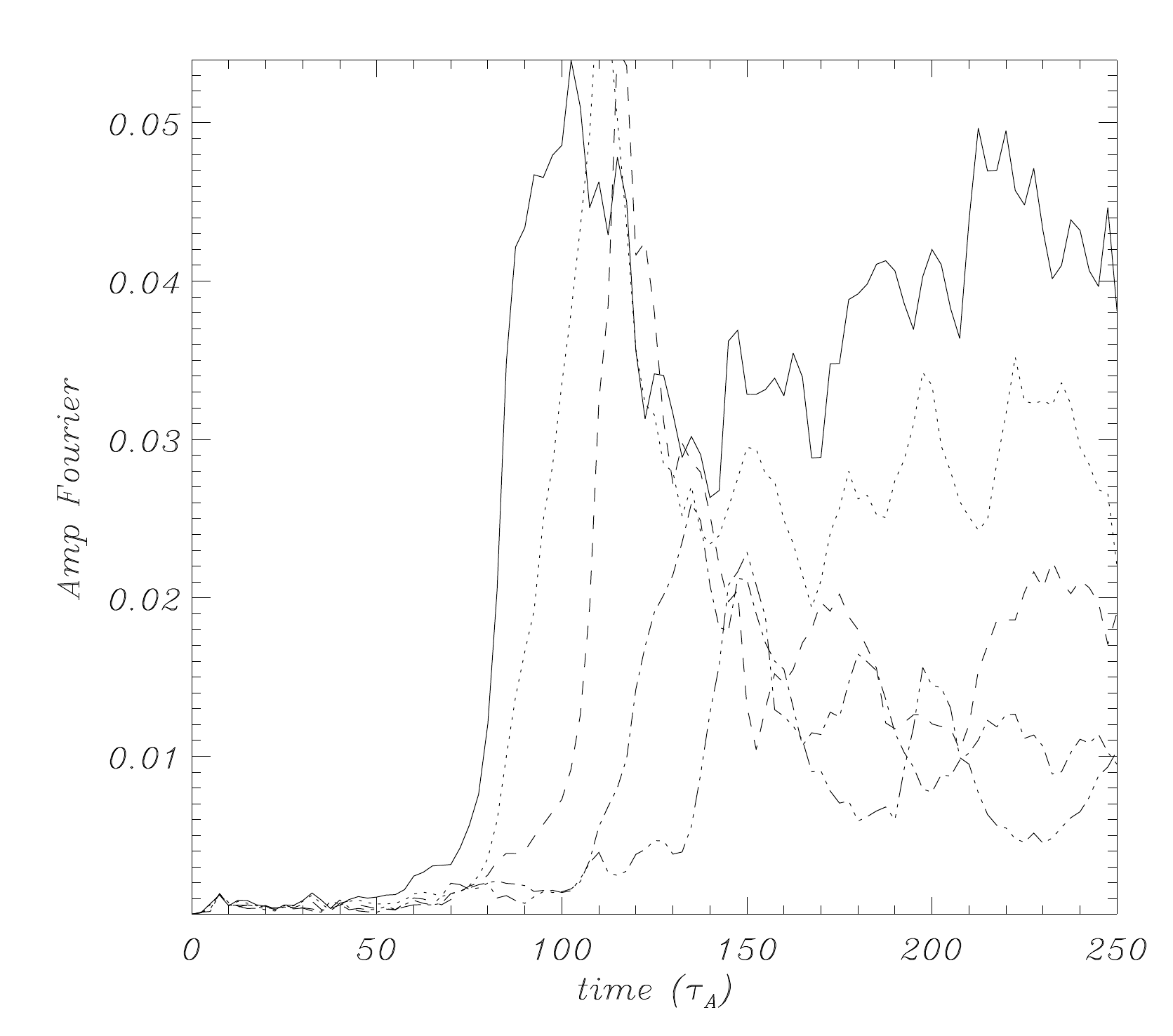}}
\caption{\small Same as in  Fig.~\ref{sumfourtwist} but in the $z-$direction  320 points, instead of 80, have been used.}\label{sumfourtwisthres}
\end{figure}

\subsubsection{Effect of numerical resolution}

In this section about twist we have used the intermediate resolution, i.e., 
[400, 400, 80]. However, the nature of the oscillations in this case is not as
simple as in the untwisted case, characterized by a dominant large wavelength
along the tube. Twist introduces helical motions that produce small wavelengths
along the tube and for this reason it is interesting to check the effect an
increased resolution on the instability onset times. We have repeated the same
simulations as in  Fig.~\ref{sumfourtwist} but using [400, 400, 320] points. The
results are shown in Fig.~\ref{sumfourtwisthres} and indicate that for the values
of twist of $B_\phi/B_z=0,0.1, 0.2$ the onset times are essentially the same as
in the case  [400, 400, 80] (compare with Fig.~\ref{sumfourtwist}). We already
found this behavior in the untwisted case (see Section~\ref{sectresol}). The
differences arise in the situation of strong twist, $B_\phi/B_z=0.3$ and $0.4$
indicating that insufficient resolution along the tube delays the appearance of
the instability. Therefore, increasing the resolution in the $z-$direction
slightly reduces the slopes of the curves in Figs.~\ref{tauinst} and
\ref{tauinsttaud}. 

\begin{figure}[!hh] \center{\includegraphics[width=7cm]{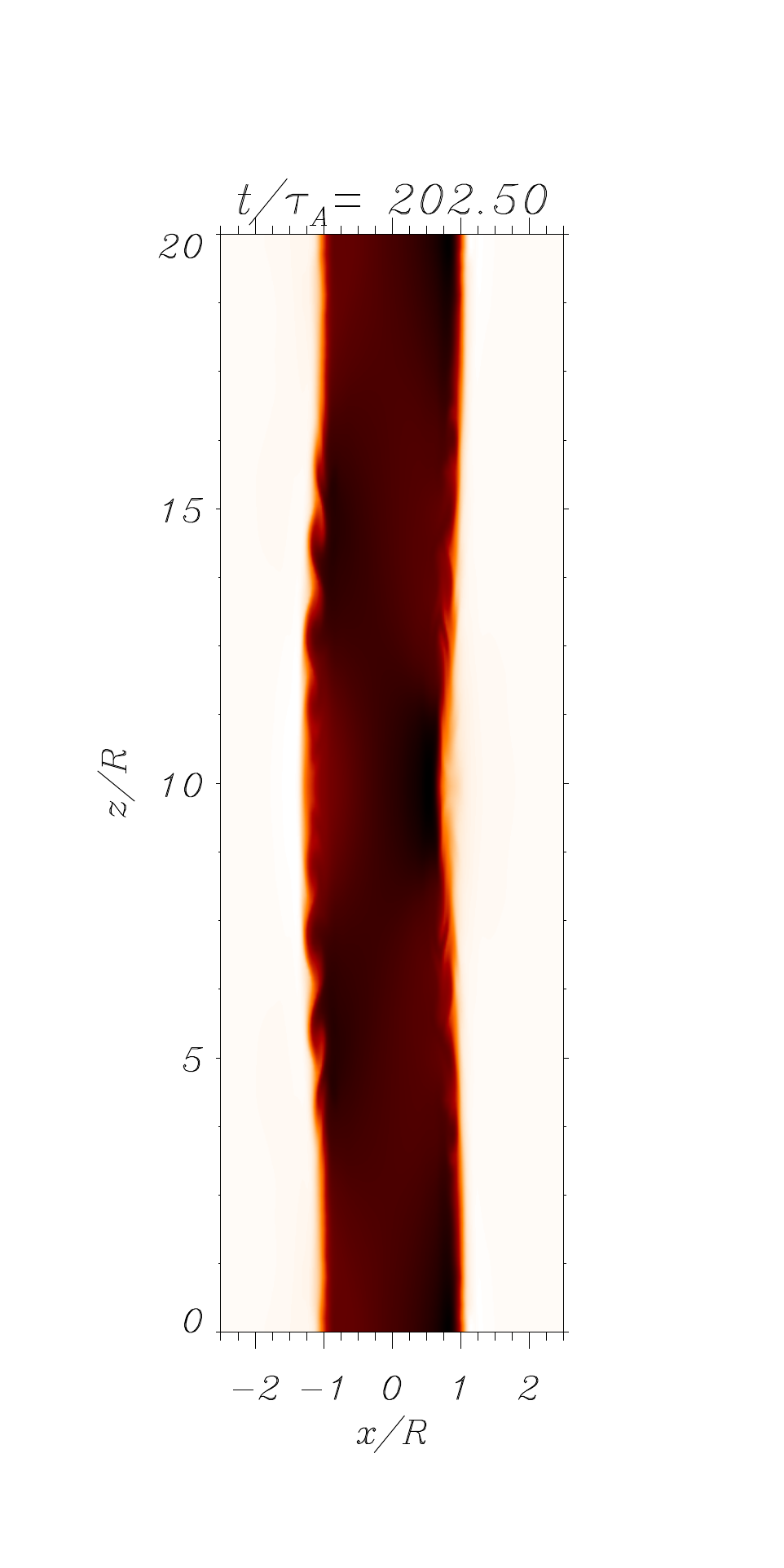}}
\caption{\small Snapshot of the two-dimensional density distribution (at the
plane $y=0$) at a given time. In this case $l/R=0.3$ and
$B_\phi/B_z=0.4$.}\label{denstwistlalong} \end{figure}

Further evidence of the structuring in the $z-$direction is found in  the
longitudinal density distribution represented in Fig.~\ref{denstwistlalong}. The
vortices are present along the loop, contrary to the situation for the untwisted
case.

\section{Conclusions and Discussion}\label{conclusions}

In this work we have investigated numerically the development of the KHI at the 
edge of transversally ($m=1$) oscillating tubes. By means of numerical
simulations and specific analysis methods to interpret the contribution of the
different azimuthal wavenumbers we have extracted several conclusions about the
physics involved in this problem. In the numerical experiments, we have found a
clear excitation of the azimuthal number $p=2$, with half the period of the
transverse oscillation. This result corroborates the analytic predictions of
\citet{rudermanetal2010,rudgos2014}. However, according to our simulations the
excitation of the first fluting, $p=2$, does not produce a faster attenuation of
the $m=1$ because of the driven nature of this fluting oscillation produced by
the nonlinear terms in the MHD equations. The excitation of $p=2$ is essentially
unaltered when magnetic twist is included in the model.

Contrary to the initial predictions of \citet{soleretal2010}  about the
stabilizing effect of a weak magnetic twist using a Cartesian slab model {\bf
\citep[see also the work of][ in a cylindrical model]{temurietal2015}}, we have
found that at least for thin inhomogeneous layers  a small twist does not
significantly delay the appearance of the instability. The same results apply to
the case with a discontinuous density profile, which is always the most unstable
configuration. Under such conditions the instability might develop during the
weakly attenuated global oscillations. This agrees with the results of the
submitted paper of \citet{howsonetal2017}, in which further details about the
possible enhanced contribution to the heating due to greater Ohmic dissipation
produced by twist is investigated. However, the observations of damped coronal
loop oscillations usually show that the damping is quite strong since the
damping per period is typically in the range $1 < \tau_{\rm D}/P < 3$ \citep[for
example, TRACE oscillations have periods of the order of 2-10 minutes and short
damping times of the order of 3-20 minutes, see also the recent study
of][]{goddardnaka2016}. In any case one needs to be careful when
interpreting the damping times from observations, since these times may strongly
depend on the amplitude of the perturbation, the EUV channel in which they are
detected, the temperature difference between the inside and outside of the loop,
and also the resolution of the instrument. For instance,
\citet{antolinetal2016,antolinetal2017} describe  apparent decayless
oscillations and different damping time depending on the EUV channel and
instrument resolution.

Our study indicates that when the inhomogeneous layer that connects the interior
of the tube with the external medium, is significantly wide (i.e., thick layers)
then the effect of twist is relevant to the stabilization of the tube with
respect to the KHI. The detection of the instability in such a case is expected
to be difficult since it is triggered on a time-scale much longer than the
characteristic damping time of the global oscillation, and the loop is barely
oscillating transversally with a significant amplitude. But it may happen that a
seemingly static loop suddenly starts developing the KHI if it has been displaced
transversally before the beginning of the observations.

Better observations with high spatial resolution are still required to properly
resolve the inhomogeneous layers of oscillating magnetic tubes and to provide a
definitive answer about the role of resonant damping in the attenuation of the
oscillations and therefore in the development of the KHI and the associated
heating. In this regard, \citet{okamotoetal2015} claim to have found a compelling
signature of resonant damping on high spatial, temporal, and spectral resolution
observations of a solar prominence (a thread), and point to the relevant role of
the KHI in these observations. Along this line, \citet{kuridzeetal2015a} have
recently observed the fast disappearance of rapid redshifted and blueshifted
excursions in the H$\alpha$ line that has been interpreted as a consequence of the
heating of the structures due to the development of the KHI at their boundary
during transverse motions.

\begin{acknowledgements} J.T. acknowledges support from MINECO and UIB through a
Ram\'on y Cajal grant and also support by the Spanish MINECO and FEDER funds through
project AYA2014-54485-P. N. M. acknowledges the Fund for Scientific Research-Flanders
(FWO-Vlaanderen). T. VD. was supported by an Odysseus grant of the FWO Vlaanderen,
the IAP P7/08 CHARM (Belspo) and the GOA-2015-014 (KU~Leuven). This project has received funding from the
European Research Council (ERC) under the European Union's Horizon 2020 research and
innovation programme (grant agreement No 724326). The authors also thank R. Soler and
M. Goossens,  for their comments and suggestions that helped to improve the paper.
The support from ISSI (International Space Science Institude) through the team
``Towards Dynamic Solar Atmospheric Magneto-Seismology with New Generation
Instrumentation" lead by R. Morton and G. Verth is also acknowledged.
\end{acknowledgements}

\end{document}